\documentclass[usenatbib]{mnras} 
\pdfoutput=1
\usepackage{newtxtext,newtxmath}

\usepackage[T1]{fontenc}
\usepackage{ae,aecompl}

\usepackage{hyperref}
\hypersetup{draft}
\usepackage{amssymb,multirow,tabularx}
\usepackage{amsmath}
\usepackage{graphicx}
\usepackage{BibDef}
\usepackage{xcolor}
\usepackage[normalem]{ulem}

\newcommand{\msun}{~\rm M_{\large \odot}}
\newcommand{\ud}{~{\rm d}}
\newcommand{\udnsp}{{\rm d}}

\title[MBH triplets and LISA]{Post-Newtonian Evolution of Massive Black Hole Triplets in Galactic Nuclei: IV. Implications for LISA}

\author[Bonetti et al.]{Matteo Bonetti$^{1,2}$\thanks{E-mail: matteo.bonetti@unimib.it}, Alberto Sesana$^3$, Francesco Haardt$^{1,2}$, Enrico Barausse$^{4,5}$\& 
\newauthor Monica Colpi$^{6,2}$\\
$^1$DiSAT, Universit\`a degli Studi dell'Insubria, Via Valleggio 11, 22100 Como, Italy\\
$^2$INFN, Sezione di Milano-Bicocca, Piazza della Scienza 3, 20126 Milano, Italy\\
$^3$School of Physics and Astronomy and Institute for Gravitational Wave Astronomy, University of Birmingham, Edgbaston,\\ Birmingham B15 2TT, United Kingdom\\
$^4$CNRS, UMR 7095, Institut d'Astrophysique de Paris, 98 bis Bd Arago, 75014 Paris, France\\
$^5$Sorbonne Universit\'es, UPMC Univesit\'e Paris 6, UMR 7095, Institut d'Astrophysique de Paris, 98 bis Bd Arago, 75014 Paris, France\\
$^6$Dipartimento di Fisica ``G. Occhialini'', Universit\`a degli Studi di Milano-Bicocca, Piazza della Scienza 3, 20126 Milano, Italy
}

\date{Accepted XXX. Received YYY; in original form ZZZ}

\pubyear{2019}

\begin{document}
\label{firstpage}
\pagerange{\pageref{firstpage}--\pageref{lastpage}}
\maketitle

\begin{abstract}
	Coalescing massive black hole  binaries  (MBHBs) of $10^{4-7}\msun$, forming in the aftermath of galaxy mergers, are primary targets of the space mission LISA, the {\it Laser Interferometer Space Antenna}.
	An assessment of LISA detection prospects requires an estimate of the abundance and properties of MBHBs that form and evolve during the assembly of cosmic structures. To this aim, we employ a semi-analytic model to follow the co-evolution of MBHBs within their host galaxies. We identify three major evolutionary channels driving the binaries to coalescence: two standard paths along which the binary evolution is driven by interactions with the stellar and/or gaseous environment, and a novel channel where MBHB coalescence occurs during the interaction with a third black hole. 
	For each channel, we follow the orbital evolution of MBHBs with physically motivated models that include a self-consistent treatment of the orbital eccentricity. We find that LISA will detect between $\approx 25$ and $\approx 75$ events per year depending on the seed model.
	We show that triple-induced coalescences can range from a few detected events up to $\sim 30\%$ of the total detected mergers. Moreover, even if the standard gas/stars-driven evolutionary channels should fail and MBHBs were to stall, triple interactions would still occur as a result of the hierarchical nature of galaxy formation, resulting in about $\approx 10$ to $\approx 20$ LISA detections per year.
	Remarkably, triple interactions among the black holes can produce coalescing binaries with large eccentricities ($\gtrsim 0.9$) upon entrance into the LISA band. This eccentricity will remain significant ($\sim 0.1$) also at merger, requiring suitable templates for parameter estimation. 
\end{abstract}

\begin{keywords}
	black hole physics -- galaxies: kinematics and dynamics -- gravitation -- gravitational waves -- methods: numerical
\end{keywords} 

\section{Introduction}
\label{sec:Intro}

The existence of massive black hole binaries (MBHBs) is a key prediction of our current $\Lambda$CDM cosmological model \citep{White1978,Kauffmann2000,Volonteri2003}. The ubiquitous presence of MBHs at the centre of bright nearby galaxies \citep[see, e.g.][]{Kormendy1995,Magorrian1998,Ferrarese2000,Gebhardt2000} and the existence of luminous quasars at redshift $z$ as large as 7 \citep{Mortlock2011,Banados2018} hint at evolutionary pathways in which galactic haloes, growing through repeated mergers and accretion of cold gas along dark matter filaments, drag their central MBHs to form close pairs on kpc scales. In the interaction with the stellar and gaseous background of the new galaxy, gravitational and hydrodynamical torques control and guide the sinking of the two MBHs toward the galactic core where coalescence takes place \citep[see e.g.][for a review]{Colpi2014}.

Coalescing MBHBs forming in the aftermath of galaxy collisions are among the loudest sources of gravitational waves (GWs) in the Universe, and key targets of the ESA led space mission LISA, the {\it Laser Interferometer Space Antenna} \citep{LISA2017}. For the first time, LISA will survey the GW frequency window from about 100 $\mu$Hz to 100 mHz, where the late inspiral, merger and ring-down of MBHBs will be best observed  for massive black holes (MBHs) with masses between $10^4\msun$ and $10^7\msun.$ 
While the galaxy mass function suggests that these low mass MBHs may be  the most common, they are the least known in terms of basic demographics, birth, growth, dynamics and connection to their galaxy hosts \citep{Kormendy2013}. Nevertheless, LISA's exploration volume is immense, and will allow us to detect GW signals from MBHBs coalescing at redshifts as large as 20, when the Universe was only 180 Myrs old, down to the climax of star formation at $z\sim 2$ \citep{Madau2014rev}. By observing deep into the high-redshift Universe LISA will tell us when the first MBHBs formed and how they evolved across cosmic epochs, providing the first census of this new population of GW sources \citep{Haehnelt1994,Jaffe2003,Wyithe2003,Enoki2004,Sesana2004,Sesana2005,Jenet2005,Rhook2005,Barausse2012,Klein2016}.

However, in order to fully plan (and exploit) LISA observations, theoretical models are needed to characterise this new population of sources and make predictions for the rate of MBHB coalescences as a function of redshift. This rate depends sensitively on the MBHB mass and mass ratio, which are related to those of the interacting galaxies, and on the mechanisms leading to coalescence. Tracing the dynamics of MBHBs during a galactic merger is extremely challenging, as the process is multi-scale and depends on a variety of physical processes. Indeed, LISA MBHBs encompass a wide interval of masses, from $10^4\msun$ to a few $10^7\msun$, mass ratios $q$ from $\sim 0.1$ to $\sim 1$, and coalescence redshifts ranging from $z\sim 20$ to the present, implying very different environments and initial conditions for each binary. 
Moreover, the dynamical range is huge and MBHs have to travel long distances before coalescence, starting from the typical scale of a galactic merger (i.e. tens to hundred kpc, depending on the encounter geometry), through mpc separations (where GWs start driving the inspiral) and down to $\mu$pc scale where coalescence takes place. Inside the newly formed galaxy, gas and stars concur to extract MBH orbital energy and angular momentum, thus affecting the evolution, but
these interactions are complex because of feedback processes (from AGNs and supernovae) and the multi-phase nature of the gas component.

Furthermore, bottlenecks are possible along the way to coalescence. MBHs can wander in the outskirts of the new galaxy if the mass ratio of the progenitors is small
(below about $0.1$), as tidal stripping and ram pressure can remove stars and gas in the MBH surroundings, reducing the effectiveness of dynamical friction 
at dragging the MBHs to the nuclear region \citep{Taffoni2003,Callegari2011}. 
This can produce a population of wandering black holes on scales in excess of 10-100 pc~\citep{Tremmel2017,Dosopoulou2017}. In gas-rich environments and again on the large scales, formation of massive gas clumps (before they fragment to form stars) can lead to MBH stochastic dynamics for comparable masses, and to scattering of the MBHs off the disc plane under rather generic conditions. This may widen the distribution of MBHB coalescence times (measured as ``delays'' from the galaxy merger),
which can range from $\approx 100$ Myrs to several Gyrs. Individual mergers, which can be studied using full gas-free N-body simulations \citep{Berczik2006,Khan2011,Sesana2011b,Khan2012,Khan2012b,Khan2013,Vasiliev2014,Vasiliev2015,Vasiliev2016,Gualandris2017}, high resolution SPH simulations of galaxies in isolation \citep{Mayer2007,Dotti2007,Callegari2008,Lodato2009,Callegari2011,Fiacconi2013,Capelo2015,Roskar2015,DelValle2015,Mayer2016,Goicovic2017,Lupi2015,Tamburello2017} or in fully cosmological contexts \citep{Vogelsberger2014,Khan2016,Schaye2015,Tremmel2017,Tremmel2018}, can indeed reveal the complexity of the black hole interaction with the (large scale) environment.

On smaller scales $\sim$pc, once the MBHs have formed a binary, stellar hardening can become a major process guiding the journey to coalescence. Stars are scattered away individually from the binary, which leads to  loss-cone depletion in phase-space, and eventually binary stalling \citep{Yu2002}.
Recently, however, it has been demonstrated that regimes of full loss-cone can be attained by enhanced stellar diffusion in phase space due to the presence of triaxiality in the galaxy's gravitational potential \citep{Yu2002,Khan2011,Khan2013,Vasiliev2014,Gualandris2017} or to galactic rotation \citep{Holley-Bockelmann2015}.  
Alternatively, if a significant amount of gas is present in the form of a (massive) circumbinary disc, the evolution of the MBHB towards the GW-dominated regime may be faster, because MBHBs are expected to undergo planetary-like migration \citep{Haiman2009,Dotti2015}.   
A third process can drive a MBHB down to coalescence when the galaxy host is not isolated. Particularly at high redshifts, when galaxy mergers are commonplace, a third MBH can
approach the binary as a result of dynamical friction, forming an interacting `triplet'.
If the triplet is hierarchical, the binary may undergo the Kozai-Lidov oscillations \citep{Kozai1962,Lidov1962}, which tend to secularly increase the eccentricity of the inner binary, eventually driving it to coalescence. In addition, whenever the triplet becomes unstable, chaotic 3-body interactions may further enhance the coalescence rate \citep{Blaes2002,Hoffman2007,Amaro-Seoane2010,Kulkarni2012,Bonetti2016,Bonetti2018a,Ryu2017}.

If single detailed simulations can resolve the great complexity of the MBH interactions, semi-analytical models (SAMs) provide the large picture of MBHB coalescence \citep{Kauffmann2000,Volonteri2003, Sesana2011a,Barausse2012,Valiante2016,Ricarte2018a,Dayal2019}.
SAMs are optimally suited to perform statistical studies of the formation and
evolution of binary and multiple MBH systems. While less realistic than full-fledged simulations, SAMs are vastly superior from the
point of view of computational performance, thus allowing a systematic exploration of many different scenarios for the formation and co-evolution of MBHs within their host galaxies.  
In particular, as shown by \citet{Klein2016}, the MBHB merger rate can reduce significantly when astrophyiscally motivated ``delays'' times are accounted for in a merger scheme, also leading to a shift to lower $z$ of the maximum rate.

In this paper we combine the SAM developed by \citet{Barausse2012} (and later extensions)  with a self-consistent treatment of the dynamical interactions driving
MBHBs to coalescence. In particular, we highlight the effect and evolution of the orbital eccentricity during the standard stellar and gas-hardening phases, and also in the context of a novel coalescing channel involving the formation of MBH triplets \citep{Blaes2002,Hoffman2007,Kulkarni2012,Bonetti2018a}.
This channel arises naturally when MBHBs have a ``long'' hardening time. Under these circumstances, a subsequent galaxy merger may deliver a third MBH that is close enough to the MBHB to promote coalescence. In this context \citet{Bonetti2018a,Bonetti2018b} explored a large sample of 3-body numerical simulations, with 
the goal of implementing this triplet channel in SAMs.

Our aim is to characterise MBHB coalescences and forecast LISA detection prospects with physically motivated models for the time delays between galaxy and MBHB mergers.\footnote{The method was previously applied in the context of the nHz pulsar timing array experiments, which target GW signals and backgrounds emitted by $10^{8-9}\msun$ MBHBs at low redshifts \citep{Bonetti2018b}.}
We consider two conceptually different dynamical scenarios: a fiducial model in which MBHBs merge on timescales of millions-to-billions of years, depending on the properties of the host galaxy (and in particular on its gas and star content); and an extreme model in which all MBHBs stall at about their hardening radius, so that their coalescence is prompted only by the interaction with a third/multiple MBHs. Clearly, the latter is the most pessimistic scenario, yielding to the lowest LISA rates.

The paper is organised as follows: in Section~\ref{sec:SAM} we briefly review the main features of the SAM adopted to track structure evolution. In Section~\ref{sec:basics} we describe the physical processes that drive MBHBs to coalescence, including stellar hardening, gas-driven migration and triplet dynamics, and how we generate a mock catalogue of LISA events. In Section~\ref{sec:GW_emission_detection} we detail the formalism employed to compute waveforms and signal-to-noise ratios (hereafter S/N), and in Section~\ref{sec_results} we present our main results. Finally, Section~\ref{sec:conclusions} is devoted to a discussion of the results and to our conclusions. 

\section{Semi-analytic galaxy formation model}
\label{sec:SAM}

SAMs of galaxy and MBH formation and evolution represent a very flexible tools to explore MBH merger rates at a fairly modest computational cost when compared to the more sophisticated, but much dearer, hydrodynamical simulations \citep[e.g.][]{Springel2010,Guillet2011,Vogelsberger2014,Schaye2015,Rodriguez-Gomez2015,Kelley2017,Wadsley2017,Tremmel2018}.
For this reason, they have been extensively used, among other applications, to compute projected merger rates for MBHs \citep{Sesana2004,Sesana2005,Sesana2007b,Sesana2011a,Tamanini2016,Klein2016} and (more recently) extreme mass ratio inspirals \citep{Babak2017} in the LISA band, as well as stochastic background amplitudes and resolved event rates for pulsar timing arrays \citep{Sesana2009,Dvorkin2017,Ryu2017,Bonetti2018b}.

In this work, we utilise the SAM of \citet{Barausse2012}, with the improvements described in \citet{Sesana2014}, \citet{Antonini_Barausse2015} and \citet{Antonini2015}. 
This model, already employed in \citet{Klein2016,Dvorkin2017,Babak2017,Bonetti2018b} to explore the detection prospects of the future LISA mission and current pulsar timing array campaigns, 
evolves the baryonic galactic structures (including MBHs) along dark matter merger trees produced with an extended Press-Schechter formalism, modified to reproduce the results of N-body simulations~\citep{Press1974,Parkinson2008}. We have checked that our results are robust against the time step of the merger tree \citep[see e.g.][]{Shimizu2002}.
Besides the MBHs, the code also follows the evolution of several baryonic components.
These include for instance a hot chemically primordial intergalactic medium, and a cold chemically enriched interstellar medium, which eventually forms stars. Both the cold gas and stellar components exist as discs and/or spheroids.
On smaller scales, the model includes a pc-scale nuclear gas reservoir ~\citep{Barausse2012,Antonini_Barausse2015,Antonini2015},\footnote{If the nuclear gas reservoir is more massive than the hosted MBH, we define a galaxy as gas-rich.} whose growth rate is linearly correlated with star formation in the spheroid \citep{Granato2004,Haiman2004,Lapi2014}, which in turn takes place in our model after major galactic mergers and bar instabilities in the galactic disc. This reservoir is available to accrete onto the central MBH on the gas viscous timescale \citep{Sesana2014} and/or to form a nuclear star cluster \citep{Antonini_Barausse2015,Antonini_Barausse2015}.
Besides forming from this in-situ star formation in the nuclear gas, the nuclear star cluster 
also forms from migration of globular clusters to the galactic centre \citep{Antonini_Barausse2015,Antonini2015}.
Supernova and AGN feedback are included in the model. The latter, together with the aforementioned prescriptions for MBH accretion, ensures
that MBHs grow to reproduce the local observed correlations with host mass and velocity dispersions at $z=0$~\citep{Barausse2012,Sesana2014,Barausse2017}.

The features that most impact the MBH merger rates for LISA are the initial mass function of the MBH seeds at high redshift, and the mechanisms (with their related timescale) which drive MBHBs to coalesce after a galaxy merger. Regarding the former, we adopt two competing scenarios for the seeds. In a ``light-seed'' (LS) scenario, we make the hypothesis that MBH seeds originate from the remnants of pop III stars forming in high-redshift, low-metallicity galaxies \citep[e.g.][]{Madau2001}. To model this process, we assume that only the rarest, most massive haloes (collapsing from the $3.5\sigma$ peaks of the primordial density field
at redshift between 15 and 20) contain a MBH seed \citep{Madau2001}. The mass of
this seed is assumed to be $\sim 2/3$ of the initial pop III mass (to account for mass losses during stellar collapse), which is in turn chosen randomly from a log-normal distribution centred on 300 $M_\odot$ and with r.m.s. of 0.2 dex, with a mass gap  between 140 and 260 $\msun$, to account for the formation of pair-instability supernovae \citep{Heger2002}.

In the ``heavy-seed'' (HS) scenario, we model MBH seeds as forming from the collapse of protogalactic discs as a result of bar instabilities. Following \citet{Volonteri2008} we describe the process by a single free parameter, i.e. the critical Toomre parameter
$2\lesssim Q_c\lesssim3$, marking the onset of instability. We set $Q_c=2.5$, and we only seed haloes at $15<z<20$ as in \citet{Volonteri2008}.

As for the timescale on which MBHBs coalesce after a galaxy merger\footnote{In our SAM, galaxy mergers take place a dynamical friction time \citep{Boylan-Kolchin2008} after their host haloes merge. We also account for environmental effects on satellite haloes (tidal stripping and evaporation) following \citet{Taffoni2003}.} (i.e. the ``delays'' between galaxy and MBHB coalescences), we adopt, as in \citet{Bonetti2018b}, two different prescriptions.
In a more realistic scenario (labelled as {\it Model-delayed}) we assume that MBHBs form promptly after a galaxy merger, and that their evolution is driven by a combination of three possible processes: stellar hardening (on timescales $\lesssim$ few Gyr),
gas-induced migration (on timescales of $\sim 10^7-10^8$ yr if enough gas is available), and triple MBH interactions (on timescales of $\sim 10^8-10^9$ yr), if a third MBH is brought in by a subsequent galaxy merger.

In a more pessimistic scenario ({\it Model-stalled}), we instead make the assumption that all
MBHBs stall, such that coalescences can only occur because of the formation of a MBH triplet, which can promote binary mergers via either Kozai-Lidov resonances \citep{Kozai1962,Lidov1962}
or chaotic three-body interactions \citep{Bonetti2016,Bonetti2018a}.
The evolution of MBH triplets is modelled by fitting a set of simulations \citep{Bonetti2018a} produced with the three-body post-Newtonian code presented in \citet{Bonetti2016}. 

In summary, we consider four different evolutionary models for MBHs in galactic haloes, according to the seeding and delay prescriptions. For each model, the SAM provides us with a set of events labelled by coalescence redshift, mass of the primary MBH $m_1$, mass ratio $q$ (i.e. $q=m_2/m1\leq1$, being $m_2$ the less massive binary component), and a flag identifying the evolutionary path followed, i.e. stellar-, gas-, triplet-driven dynamics. This information is then used to construct the comoving number density of coalescing MBHBs $n$, per unit redshift and (logarithmic) unit of $m_1$ and $q$: 
\begin{equation}
\dfrac{\ud^3n}{\ud z \ud \log_{10}m_1 \ud \log_{10}q }.
\label{eq:dndz}
\end{equation}
Equation~(\ref{eq:dndz}) above can be easily converted into the number $N$ of coalescences per unit observer time as~\cite{1994MNRAS.269..199H}
\begin{align}\label{eq:rate}
&\dfrac{\ud^4N}{\ud z \ud \log_{10}m_1 \ud \log_{10}q \ud t} = \nonumber\\
&\dfrac{\ud^3n}{\ud z \ud \log_{10}m_1 \ud \log_{10}q } \times \dfrac{4\pi c \ d_L^2}{(1+z)^2},
\end{align}
where $d_L$ denotes the luminosity distance.

\section{Synthetic catalogues of coalescing MBHBs}
\label{sec:basics}

Our aim is to produce a simulated catalogue of events observable by LISA. To that purpose, we need to convert the intrinsic merger rate (equation~\ref{eq:rate}) into an instantaneous number of systems across the sky per (logarithmic) units of eccentricity $e$ and observed orbital frequency $f_{\rm orb}$. It is useful to relate the evolution of binaries in the $(\log_{10}f_{\rm orb}, \log_{10}e)$ plane as the flow of a fluid in a two dimensional space. If the flow is stationary,
the continuity equation implies that the fluid's density along a flow tube is inversely proportional to the velocity $v$ and to the flow tube's cross section $\Sigma$. 
We can thus write
\begin{align}\label{eq:5Ddistribution}
&\dfrac{\ud^5N}{\ud z \ud \log_{10}m_1 \ud \log_{10}q \ud \log_{10}f_{\rm orb} \ud \log_{10}e} = \nonumber \\ 
& \dfrac{\ud^4N}{\ud z \ud \log_{10}m_1 \ud 
	\log_{10}q \ud t} \times \frac{1}{\Sigma v}\,,   
\end{align}
where ${1}/(\Sigma v)=\tau \times \udnsp p/(\udnsp \log_{10}f_{\rm orb} \udnsp \log_{10}e)$ is the product of the average binary lifetime $\tau$ and the probability density $\ud^2 p/(\udnsp \log_{10}f_{\rm orb} \udnsp \log_{10}e)$ that a binary is observed at a given $f_{\rm orb}$ and $e$. The practical computation of this quantity is described in Sec. \ref{sec:catalog_comp}. The normalisation ${\udnsp^4N}/(\udnsp z \ud \log_{10}m_1 \udnsp \log_{10}q \,\udnsp t)$ is provided by our SAM. Note that equation~(\ref{eq:5Ddistribution}) generalises the treatment of \citet{Sesana_Vecchio2008} to binaries with eccentricity.

The SAM informs us of the type of evolutionary path (stellar-, gas- or triplet-driven) followed by binaries, without any information about the time evolution of the orbital elements. Such details are however necessary in order to compute $1/(\Sigma v)$ in the specific environment of the galaxy merger remnant. As already discussed, we consider two ``standard'' evolutionary channels, i.e. stellar-driven dynamics and gas-driven dynamics \citep[c.f. also][]{Antonini2015,Klein2016}, and a third, less standard dynamical process involving the formation and evolution of triple systems of MBHs \citep{Bonetti2016,Bonetti2018a,Bonetti2018b}. All such paths are, in principle, followed by a final, short phase dominated by GW energy losses. In the following, we will discuss in detail our implementation of MBHB dynamics.

\subsection{Dynamics of MBHBs}
\label{sec:channels}

\subsubsection{GW-driven dynamics}
Irrespective of the actual processes involved in the early stages of the orbital evolution of a MBHB, the last stages will always be dominated by GW radiation losses. For the first two evolutionary channels mentioned above (i.e. stellar and gas driven dynamics), we model the GW phase adopting the orbital averaged equations derived by \citet{Peters1964}, which describes the time evolution of the orbital frequency $f_{\rm orb}$ and eccentricity $e$:
\begin{align}\label{eq:dfdt_gw}
\frac{\ud f_{\rm orb}}{\ud t}\Big{|}_{\rm gw}=\frac{96 G^{5/3}}{5 c^5} (2\pi)^{8/3} \mathcal{M}^{5/3}\, f_{\rm orb}^{11/3}\, {\cal F}(e),\\
\frac{\ud e}{\ud t}\Big{|}_{\rm gw}=-\frac{G^{5/3}}{15 c^5} (2\pi)^{8/3} \mathcal{M}^{5/3}\, f_{\rm orb}^{8/3}\, {\cal G}(e),
\label{eq:dedt_gw}
\end{align}
where $\mathcal{M}$ is the binary source-frame chirp mass, defined as
\begin{equation}
\mathcal{M} = (m_1 m_2)^{3/5} (m_1+m_2)^{-1/5}
\end{equation}
while ${\cal F}(e)$ and ${\cal G}(e)$ are algebraic functions of the eccentricity:
\begin{align}\label{eq:peters}
{\cal F}(e) &= \dfrac{1+73/24 e^2 + 37/96 e^4}{(1-e^2)^{7/2}},
\\
{\cal G}(e) &= \dfrac{304 e + 121 e^3}{(1-e^2)^{5/2}}.
\end{align}
In the case of triple MBH interactions, we instead adopt the more refined treatment of \citet{Bonetti2016}, briefly described in the next Section~\ref{sec:triple}.  

\subsubsection{Stellar-driven dynamics}  
\label{sec:stellar_driven}

In the aftermath of a merger involving two gas-poor galaxies, the main process driving the early evolution of a MBHB is the scattering of background stars \citep{Mikkola1992,Quinlan1996,Sesana2006}. Stars intersecting the MBHB orbit can extract a non-negligible fraction of the binary energy and angular momentum because of the slingshot mechanism. The secular, cumulative effect of a large number of 3-body encounters is effective at hardening the MBHB. Eventually, when the orbital semi-major axis shrinks enough, GW emission takes over and the binary quickly evolves to coalescence. The whole process can be described by the relatively simple, albeit approximate, coupled differential equations: 
\begin{align}\label{eq:dfdt_stargw}
\frac{\ud f_{\rm orb}}{\ud t}\Big{|}_{\rm star}&= \frac{3 G^{4/3}}{2 (2\pi)^{2/3}} \frac{H\rho_i}{\sigma} M^{1/3}\, f_{\rm orb}^{1/3},\\
\frac{\ud e}{\ud t}\Big{|}_{\rm star}&= \frac{G^{4/3} M^{1/3} \rho_i H K}{(2\pi)^{2/3}\sigma}\, f_{\rm orb}^{-2/3},
\label{eq:dedt_stargw}
\end{align}
where $M=m_1+m_2$ is the binary total mass, $\sigma$ is the velocity dispersion of the host galaxy, $\rho_i$ is the stellar density at the binary influence radius (i.e. the  radius containing twice the MBHB mass in stars), whereas $H$ and $K$ represent numerical factor calibrated against scattering experiments \citep[see][for complete details]{Quinlan1996,Sesana2006}. The efficiency of the stellar hardening saturates at the hardening radius, $a_h\sim Gm_2/4\sigma^2$ \citep[where $m_2$ is the mass of the secondary,][]{Quinlan1996}, and beyond that point the MBHB hardens at a constant rate. 

In order to make use of equations~(\ref{eq:dfdt_stargw}-\ref{eq:dedt_stargw}), we need to link the properties of the stellar distribution to the mass of the hosted MBHB, i.e. we need to evaluate $\rho_i$ and $\sigma$ for a given $M$. These quantities can be inferred once the stellar mass $M_\star$ has been specified. For a given $M$, we numerically sample $M_\star$ from the stellar mass distribution provided by the SAM,\footnote{This procedure is needed since the evaluation of $1/(\Sigma v)$ in equation~\ref{eq:5Ddistribution} is associated with discrete values of $m_1$ and $q$ (as detailed in Section~\ref{sec:catalog_comp}).}  and we then evaluate an effective radius for the stellar spheroid with the following scaling relation \citep[see][]{Shen2003}:
\begin{equation}\label{eq:Reff_shen}
\log_{10}\left(\dfrac{R_{\rm eff}}{{\rm kpc}}\right) =
\begin{cases}
-5.54+0.56\log_{10}\left(\dfrac{M_\star}{\msun}\right)& \text{if } M_\star > 10^{10.3}\msun, \\
\ \\
-1.21+0.14\log_{10}\left(\dfrac{M_\star}{\msun}\right)& \text{if } M_\star \leq 10^{10.3}\msun.
\end{cases}
\end{equation}
Next, we model the stellar density profile of the spheroid by assuming a Dehnen profile \citep[being $\gamma$ the inner logarithmic slope; see][]{Dehnen1993}, i.e.
\begin{equation}\label{eq:dehnen_density}
\rho(r) = \dfrac{(3-\gamma)M_\star}{4 \pi}\dfrac{r_0}{r^\gamma (r+r_0)^{4-\gamma}},
\end{equation}
where the scale radius $r_0$ is inferred from the galaxy effective radius $R_{\rm eff}$ as in \citet{Dehnen1993}, 
\begin{equation}\label{eq:R_eff}
r_0\approx 1.3 R_{\rm eff} [2^{1/(3-\gamma)}-1].
\end{equation}
The density at the influence radius $\rho_i$ is then readily obtained from equation~(\ref{eq:dehnen_density}) by setting $r = r_{\rm inf}$. 
Finally, the velocity dispersion $\sigma$ is evaluated (under the assumption that the virial theorem holds) from
\begin{equation}
	\sigma \simeq 0.4 \sqrt{\dfrac{G M_\star}{r_0}},
\end{equation}
where the numerical pre-factor takes into account possible anisotropies in the stellar phase space \citep[see e.g.][]{Baes2002}.

Throughout the paper we consider two fairly different density profiles, characterised by different values of the inner logarithmic slope in equation~\ref{eq:dehnen_density}: $\gamma=1$, which gives relatively low central stellar densities, resulting in a rather weak coupling between the MBHB and its stellar environment; and $\gamma=1.75$, which instead gives much higher central densities, resulting in much stronger coupling. We initialise the MBHB at a separation such that the enclosed mass in stars matches the mass of the secondary MBH, and at this separation we draw a random eccentricity from a thermal distribution ($dp/de = 2e$). We finally evolve the binary until coalescence, using equations~(\ref{eq:dfdt_gw}-\ref{eq:dedt_gw}) and equations~(\ref{eq:dfdt_stargw}-\ref{eq:dedt_stargw}).\footnote{We assume that the two MBHs merge when they reach a separation lower than six gravitational radii, i.e. $a < 6 G M/c^2$.}

\subsubsection{Gas-driven dynamics}
\label{sec:gas_driven}

In a gas-rich galaxy merger, we model the early evolution of the MBHB adopting a simplified description of type II migration in circumbinary discs \citep{Artymowicz1994,Gould2000,Haiman2009,Roedig2011,delValle2012,Kocsis2012,Dotti2015}. In order to gain insight into the orbital evolution of the binary, we consider two different models, i.e. the fairly sophisticated description of type II migration proposed by \citet{Haiman2009} (HKM hereinafter), and the simpler
model considered by \citet{Dotti2015} (DMM herein after). 

In the HKM model, the binary is embedded in a thin circumbinary disc, with the plane of the disc aligned with the binary orbit. The orbital decay is dominated by viscous angular momentum exchange with the disc. In order to describe the evolution quantitatively, HKM assumed that the circumbinary gas forms a standard Shakura \& Sunyaev accretion disc \citep[][]{Shakura1973}. HKM found that just before the transition to GW-driven evolution, the viscous orbital decay is in the “secondary-dominated” Type-II migration regime (i.e. the mass of the secondary is larger than the enclosed disc mass). This is slower than disc-dominated Type-I migration.
For the HKM model, the orbital frequency evolves as 
\begin{align}\label{eq:HKM}
\frac{\ud f_{\rm orb}}{\ud t}\Big{|}_{\rm gas} = \dfrac{3}{2} \mathcal{D} & (2\pi)^{7/12} G^{-7/24} M^{-1/6} \left(\dfrac{\dot{m}}{\epsilon}\right)^{5/8}  \nonumber\\
& \times \left[ \dfrac{4q}{(1+q)^2} \right]^{-3/8}\alpha_{\nu}^{1/2} \delta^{-7/8} f_{\rm orb}^{19/12},
\end{align}
where $\mathcal{D} = 4.5\times 10^{-43} {\rm m^{7/8} kg^{-1/8} s^{-1}}$ is a constant that depends on the details of the model. The numerical value reported here is specific of a $\beta-$disc with fiducial values of the involved parameters We further use $\alpha_\nu=0.3$ for the disc viscosity parameter and $\epsilon=0.1$ for the radiative efficiency \citep[see][for full details]{Haiman2009}. The accretion rate in Eddington units $\dot m\equiv \epsilon \dot M c^2/L_{\rm Edd}$ is modelled as in Appendix A of \citet{Tamanini2016}, where we curbed the accretion to Eddington. Finally, $\delta=(1+q)(1+e)$ is the size of the gap \citep[in units of the binary semi-major axis, see e.g.][]{Artymowicz1994}. 

DMM give an order of magnitude estimate of the binary coalescence time-scale, proposing a simple model for the interaction between a MBHB and a corotating circumbinary accretion disc. Under this configuration, the gravitational torque exerted by the MBHB stops the gas inflow at a separation of $\delta \sim 2$ \citep[see, e.g.][]{Artymowicz1994}. At this radius the gravitational torque between the binary and the disc is perfectly balanced by the torques that determine the radial gas inflow on large scales. The analogous to equation~(\ref{eq:HKM}) for the simpler DMM model is 
\begin{align}\label{eq:DMM}
\frac{\ud f_{\rm orb}}{\ud t}\Big{|}_{\rm gas}= \dfrac{3 \dot{M}}{\mu} \sqrt{\dfrac{\delta}{1-e^2}}\,\, f_{\rm orb},
\end{align}
where $\dot{M}$ represents here the accretion rate within the disc, and $\mu$ the reduced mass of the binary. In general, the evolution implied by this last equation is somewhat faster compared to equation~(\ref{eq:HKM}). An equal mass binary with $M=2\times 10^6$ $M_{\odot}$ reaches the GW dominated regime in $\simeq 1.4 \times 10^8$ yrs, compared to $\simeq 1.8\times 10^8$ yrs in the HKM case. 

Note that the two considered models describe the orbital evolution as a change in the orbital frequency only, with the eccentricity unaffected. Indeed, as suggested by e.g. \citet{Roedig2011}, during type II migration the binary eccentricity converges to a specific value, $e_{\rm lim}\sim 0.7$, with only a mild dependence on the mass ratio. The value of the limiting eccentricity is obtained in \citet{Roedig2011} by balancing the disc angular frequency with that of the secondary at apocentre. This leads to an expression that relates $e_{\rm lim}$ to the size of the gap (in units of the binary semi-major axis) $\delta$: 
\begin{equation}
\delta^3 = \dfrac{(1+e_{\rm lim})^3}{1-e_{\rm lim}},
\end{equation} 
which, once solved for $e_{\rm lim}$, yields the limiting eccentricity. We then initialise the initial MBHB eccentricity by sampling from a Gaussian distribution with mean equal $e_{\rm lim}$ and standard deviation 0.05. We note, however, that the particular adopted ansatz does not have a strong influence on the properties of the observed LISA events, as all gas-driven mergers will present a negligible residual eccentricity (see next Section~\ref{sec:results}).

\subsubsection{Triple MBH dynamics}
\label{sec:triple}
In the two previous subsections we have described the binary orbital evolution in the cases of dry and wet galaxy mergers, respectively. In gas-rich galaxies, the actual efficiency of gas-MBHB interaction in realistic physical conditions is poorly known, and stalling of the binary is still a possibility \citep[see, e.g.][]{Lodato2009}. In dry galaxy mergers, we have seen that the efficiency of the stellar hardening saturates at the hardening radius and, unless new stars are continuously supplied to the otherwise depleting loss-cone, the whole process is fated to stop. Because typically $a_h\sim 0.1$ pc for $\sim 10^6 \msun$ black holes, it is not guaranteed that the MBHB can eventually close the gap down to separations $a_{\rm gw}\sim 10^{-3}$ pc, i.e. the separations at which GW emission alone can drive the two MBHs to coalesce within a Hubble time. 

A possible mechanism that could solve this problem is provided by triple MBH interactions. Triple MBH systems can form when a MBHB stalled at separations $\lesssim a_{\rm h}$ interacts with a third MBH carried by a new galaxy merger \citep[see, e.g.][]{Mikkola1990,Heinamaki2001,Blaes2002,Hoffman2007,Kulkarni2012}. The net outcome of a series of complex three-body interactions can effectively increase the binary eccentricity, reaching values close to unity and leading to copious GW losses, hence driving MBHBs to coalescence. A thorough assessment of the process requires a detailed description of the three-body dynamics, which we have carried out in a series of papers \citep{Bonetti2016,Bonetti2018a,Bonetti2018b}. Here we summarise the main features and results of that work.

We numerically integrate the orbits of MBH triplets formed by a stalled MBHB at the centre of a stellar spherical potential, and by a third MBH approaching the system from larger distances. The employed numerical scheme directly integrates the three-body (Hamiltonian) equations of motion up to 2.5PN order, introducing velocity-dependant forces to account for the dynamical friction on the intruder during its initial orbital decay toward the galactic centre, and for the stellar hardening \citep{Quinlan1996} of the outer binary as described in Section~\ref{sec:stellar_driven}. 

We found that over a large sample of $\sim$ 15000 simulated systems, a fraction of about 30\% experiences the merger of any one pair of MBHs in the triplet. The typical coalescence timescale is $\sim$~a few~$\times 10^{8}$ yrs, a timescale dominated by the orbital decay of the third MBH. Finally, extremely important for the study we are presenting here is the finding that coalescing binaries are generally driven to eccentricities, in excess of 0.9 and up to 0.9999 in some cases. Moreover, even at coalescence, binaries can retain a non-zero eccentricity, up to 0.1. Binaries driven by triple MBH interactions can therefore have large eccentricities when entering the LISA band. 

\begin{figure*}
	\includegraphics[scale=0.29,clip=true]{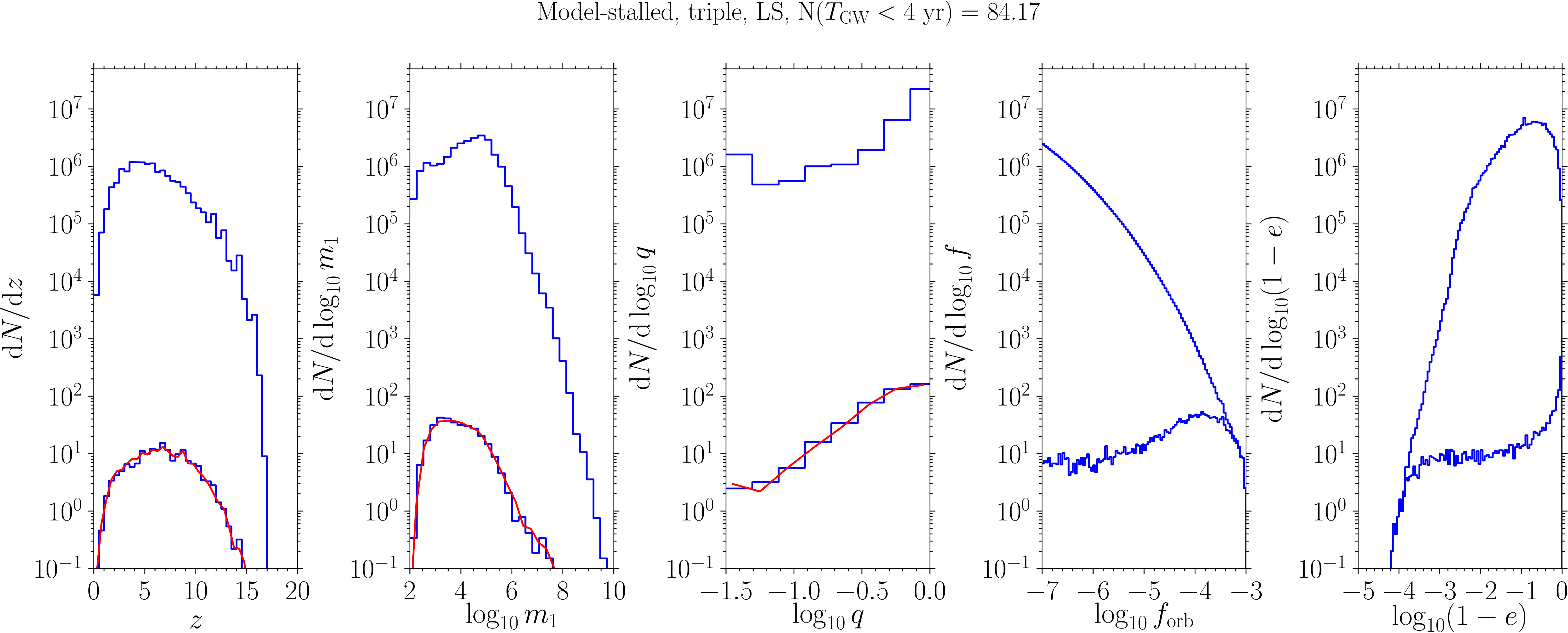}
	\caption{Monte Carlo sampling of the 5-dimensional distribution given by equation~(\ref{eq:5Ddistribution}) in the case of {\it Model-stalled}, for the light seed scenario (LS). Each panel shows the marginalised distributions over various combinations of 4 parameters. With the upper blue histogram we represent the distribution of the total number of binaries in the Universe, while the lower blue histogram is a sub-sample of those MBHBs that coalescence within 4 years (i.e. within the nominal LISA mission duration). Note the long tails in the last two panels, in particular that in the rightmost panel, showing that at the beginning of the mission there exists a population of highly eccentric MBHBs bound to merge in less than 4 years. Finally, as a consistency check of our catalogue assembling procedure, we plot as a red line the distribution given by equation~\ref{eq:rate} multiplied by 4 years.} 
	\label{fig:sampling_stalledLS}
\end{figure*}

\subsection{Catalogue compilation}
\label{sec:catalog_comp}

The quantity ${1}/(\Sigma v)=\tau \times \udnsp^2 p/(\udnsp \log_{10}f_{\rm orb} \udnsp \log_{10}e)$ appearing in equation~(\ref{eq:5Ddistribution}) is evaluated numerically following the evolution of a sample of binaries over the 2-dimensional space $(\log_{10}f_{\rm orb} ,\log_{10}e)$, covered by a grid in the range $-14\leq \log_{10}f_{\rm orb}  \leq -1$ and $-5 \leq \log_{10}(1-e) \leq 0$. Binaries are selected according to a combination of primary mass and mass ratio. Typically, for each pair $(m_1,q)$, we simulate from few tens to few thousands systems, according to the evolutionary channel followed as described in
Section~\ref{sec:channels}. In practice, for a specific binary characterised by $(m_1,q)$, we evaluate the time that the system spends in each bin of the 2-dimensional grid of frequency and eccentricity. We then sum 
over all binaries, and divide by the total number of simulated systems and by the bin area to compute ${1}/(\Sigma v)$ \citep[see Fig.~9 in][for a concrete example.]{Bonetti2018a}.

Our binaries are simulated for 45 combinations of primary mass and mass ratio, i.e. $\log_{10} m_1=2,3,4,5,6,7,8,9,10$ and $\log_{10} q = -1.5,-1.0,-0.5,-0.25,0.0$, resulting in a coarse coverage of the parameter space. Given the strong dependence of the GW losses on the binary parameters, we would need a much finer sampling of the $(m_1,q)$ plane. In order to 
interpolate in the $(m_1,q)$ plane, we
exploit the fact that the evolution of GW dominated binaries, which is described by equations~(\ref{eq:dfdt_gw}-\ref{eq:dedt_gw}), depends in a simple way on $m_1$ and $q$. Indeed,
one can rewrite equations~(\ref{eq:dfdt_gw}-\ref{eq:dedt_gw}) in terms of the logarithm of the dimensionless frequency $\tilde{f}_{\rm orb}=f_{\rm orb}GM/c^3$ and in terms of the dimensionless time $\tilde{t}=t \mathcal{M}^{5/3} M^{-8/3} c^3/G=t \eta/(G M/c^3)$, where
$\eta$ is the symmetric mass ratio $\eta=m_1\,m_2/M^2$. In the variables ($\log_{10} \tilde{f}_{\rm orb}, e, \tilde{t}$) the dependence
on $m_1$ and $q$ disappears (as expected as general relativity has no preferred scale). This allows us to interpolate our results for ${1}/(\Sigma v)$ in the $(m_1,q)$ space employing a bilinear interpolation among the four closest pairs of $(m_1,q)$.

It is important to stress again that the above procedure is valid only when the binary evolution in the $(\log_{10}f_{\rm orb},\log_{10}e)$ plane is driven by GW losses. As discussed in the previous section, before the GW-dominated regime several astrophysical processes other than GW emission drive the binary evolution. In these earlier evolutionary phases our interpolation scheme is indeed not correct. However, GWs always dominate the dynamics of MBHBs in the LISA frequency band, hence our procedure is justified.

In order to create a catalogue of LISA events, we now need to sample the distribution given by equation~(\ref{eq:5Ddistribution}). Using a standard rejection technique, we randomly generate a number of MBHBs equal to the normalization of our sampled distribution, and average over 100 realizations. In order to speed up the computation, we limit the sampling to a given minimum value of the (observed) orbital frequency. Equation~(\ref{eq:5Ddistribution}) effectively represents a snapshot of all binaries in our synthetic Universe with observed orbital frequency larger than the chosen minimum threshold (i.e. 10-100 nHz). Then, the binaries are evolved under the effect of GW emission for the nominal LISA mission lifetime of 4 years (observer time). For each system we finally compute the S/N, as detailed in Section~\ref{sec:SNR}.

As an example, we report in Fig.~\ref{fig:sampling_stalledLS} the sampling for the triplet channel of {\it Model-stalled} with the LS prescription. In each panel, the upper blue histogram represents the marginalisation of the 5-dimensional distribution (equation~\ref{eq:5Ddistribution}) over the remaining 4 parameters. The lower blue histogram is the sub-sample of MBHBs that are bound to coalesce in 4 years. The red line in the first three leftmost panels shows the same distribution but obtained directly by multiplying the merger rate (equation~\ref{eq:rate}) for 4 years; as expected, this distribution matches the randomly generated sub-sample.  

It is interesting to note that the circularity distribution (where the circularity is defined as $1-e$, rightmost panel of the figure) extends up to very low values, i.e. to eccentricities very close to unity. The implication of highly eccentric binaries for the LISA event rate will be discussed in the next sections.

\section{Gravitational wave emission and detection}
\label{sec:GW_emission_detection}

The synthetic catalogue of binaries observable by LISA described in previous sections is then used to characterise the shape and strength of the emitted GW signal. After a brief review of the LISA sensitivity, we describe the formalism employed to compute waveforms in the various stages of the MBHB orbital evolution. 

\subsection{LISA sensitivity}
\label{sec:LISA_sens}

Throughout the paper, we adopt a six-link LISA configuration (i.e. one consisting of two independent detectors). For each independent detector, we adopt the sky-averaged, single detector LISA sensitivity curve (updated to February 2018), as detailed in the ``LISA Strain Curves'' document (referenced as LISA-LCST-SGS-TN-001\footnote{See also \url{https://atrium.in2p3.fr/nuxeo/nxpath/default/Atrium/sections/Public/LISA@view_documents?tabIds=\%3A}}). Besides the instrumental noise we also take into account the effect of a large population of unresolved galactic compact binaries (mostly white-dwarf binaries) as given by \citet{Cornish2018}.
The stochastic background from that population produces a ``confusion noise'' that degrades the instrumental sensitivity at frequencies below $\sim 1$ mHz. Combining the instrumental and confusion noise contributions, the total LISA sensitivity as a function of the GW frequency $f$ is
\begin{equation}
S_n(f) = \dfrac{10}{3 L^2} \left(P_{\rm OMS} + \dfrac{4 P_{\rm acc}}{(2 \pi f)^4}\right) \left(1+\dfrac{6}{10}\left(\dfrac{f}{f_{\star}}\right)^2 \right) + S_c(f),
\end{equation}
where $L = 2.5$ Gm is the LISA arm-length, $f_{\star} = 19.09$ mHz, with
\begin{equation}
P_{\rm OMS} = 2.25 \times 10^{-22} \left[ 1+ \left(\dfrac{2{\rm mHz}}{f}\right)^4 \right] {\rm m^2 Hz^{-1}},
\end{equation}
\begin{align}
P_{\rm acc} = 9 \times 10^{-30} \Biggl[ 1+ &\left(\dfrac{0.4 {\rm mHz}}{f}\Biggr)^2 \right]\times \nonumber\\
& \times \left[ 1+ \left(\dfrac{f}{{\rm 8 mHz}}\right)^4 \right] {\rm m^2 Hz^{3}},
\end{align}
and
\begin{equation}\label{eq:Sc}
S_c(f) = A f^{-7/3} e^{-f^\alpha + \beta f \sin{\kappa f}} \left[1+ \tanh(\gamma(f_k-f))\right] {\rm Hz^{-1}}.
\end{equation}
The last term, $S_c(f)$, is the aforementioned confusion noise.
In equation~(\ref{eq:Sc}), for a 4-year mission, the parameters $A, \alpha, \beta, \kappa, \gamma$ and $f_k$ take the following values: 
\begin{align}
A &= 9 \times 10^{-45}\nonumber\\
\alpha &= 0.138\nonumber\\
\beta &= -221\nonumber\\
\kappa &= 521\nonumber\\
\gamma &= 1680\nonumber\\
f_k &= 0.00113 \rm ~Hz.
\end{align}

\subsection{Waveforms and S/N calculation}
\label{sec:SNR}

As already discussed, in the LISA frequency domain ($10^{-4}\lesssim f \lesssim 1$ Hz) the dynamics is driven by GW losses. In this section we briefly describe the formalism adopted to compute the S/N in the inspiral, merger and ringdown phases. Because, as we will see in Section~\ref{sec_results}, triple MBH interactions can lead binaries to develop large eccentricities in the LISA band, we will focus in particular on including orbital eccentricity in the waveforms. Inspiral-merger-ringdown waveforms including eccentricity typically assume small to moderate eccentricities \citep{Huerta2017,Huerta2018}, which is not necessarily adequate for our purposes (c.f. Fig.~\ref{fig:e_distr}). For this reason, we have decided to rather opt for a simplified model accounting for (arbitrarily large) eccentricity in the inspiral phase at leading Newtonian order, while neglecting higher order PN corrections as well as MBH spins. In more detail, we use the Newtonian fluxes from \citet{Peters1964} to evolve the semi-major axis and eccentricity of a given MBHB, while the S/N (of the inspiral phase) is computed as in \citet{Barack2004} by converting the Newtonian fluxes (in the various harmonics) into characteristic GW amplitudes. To further account for the effect of the merger-ringdown phase on the S/N, we also augment the inspiral S/N by a merger-ringdown contribution, which we compute from the post-inspiral portion of PhenomC waveforms. In the following, we will describe our procedure in more detail.

\subsubsection{Inspiral}
\label{sec:snr_insp}

\begin{figure}
	\includegraphics[scale=0.37,clip=true]{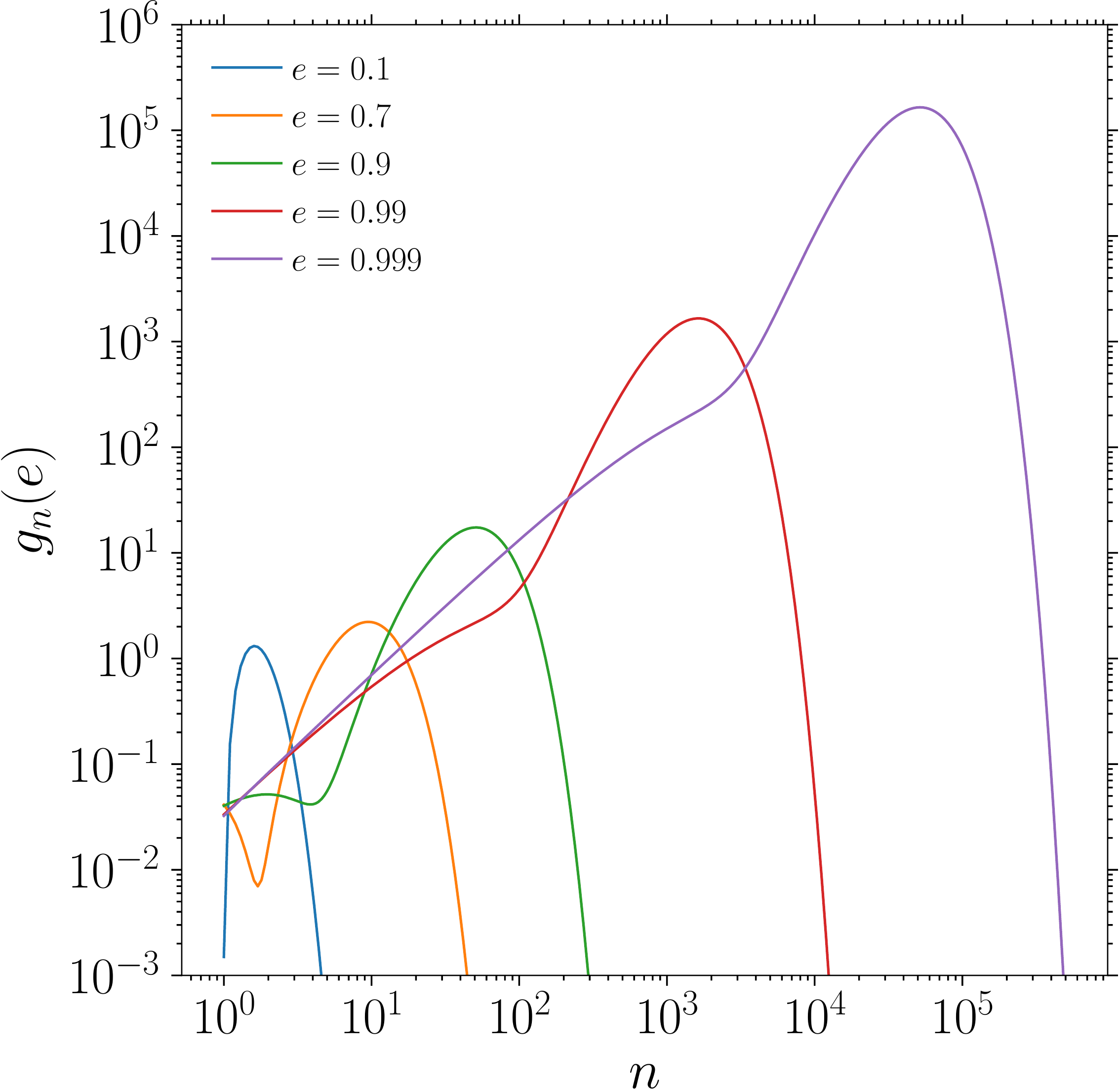}
	\caption{Examples of function $g_n(e)$ (see equations~\ref{eq:Edot_n}-\ref{eq:g_n_e}) as a function of harmonic $n$ for five values of the eccentricity, as labelled. Note how the function peak shifts at higher $n$ when $e$ approaches unity, i.e. most of the power in GW is emitted at the highest frequencies.} 
	\label{fig:gfunc}
\end{figure}

In the inspiral phase, we evolve MBHBs in the GW regime with the standard orbit-averaged dynamics given by equations~(\ref{eq:dfdt_gw}-\ref{eq:dedt_gw}) \citep{Peters1964}. The computation of the S/N requires knowledge of the emitted waveform. Gravitational radiation emitted by an eccentric binary requires a more complicated treatment compared to the standard case of a circular orbit. While for circular binaries most of the GW power is contained in the dominant quadrupolar mode, whose frequency is twice the orbital frequency, several harmonics are 
excited with comparable amplitudes in the eccentric case, i.e. the GW spectrum contains several dominant frequencies $f_n = n f_{\rm orb}$, where $n$ is the harmonic number. Therefore the total emitted power in GW is spread on a broad spectrum of frequencies, with the fraction of power per harmonic given by \citet{Peters1963}
\begin{equation}\label{eq:Edot_n}
\dot{E_n} = \dfrac{32 G^{7/3}}{5 c^5} (2\pi \mathcal{M} f_{\rm orb})^{10/3} g_n(e),
\end{equation}
where the dimensionless function $g_n(e)$ (see Fig.~\ref{fig:gfunc}) is related to the power carried by each harmonic, and is given by
\begin{align}\label{eq:g_n_e}
g_n(e) &= \frac{n^4}{32} \Bigg[\bigg(J_{n-2}(ne)-2eJ_{n-1}(ne)+\frac{2}{n}J_n(ne)\nonumber\\
&+2eJ_{n+1}(ne)-J_{n+2}(ne)\bigg)^2 \nonumber\\ &+(1-e^2)\Big(J_{n-2}(ne)-2J_n(ne)+J_{n+2}(ne)\Big)^2 \nonumber\\
&+ \frac{4}{3n^2} J_n^2(ne)\Bigg].
\end{align}
Here $J_n$ represents the $n$-th order Bessel function of the first kind. Note that in the case of a circular binary $g_n(0)=\delta_{2n}$, where $\delta_{mn}$ is the standard Kronecker delta.

\begin{figure}
	\includegraphics[scale=0.37,clip=true]{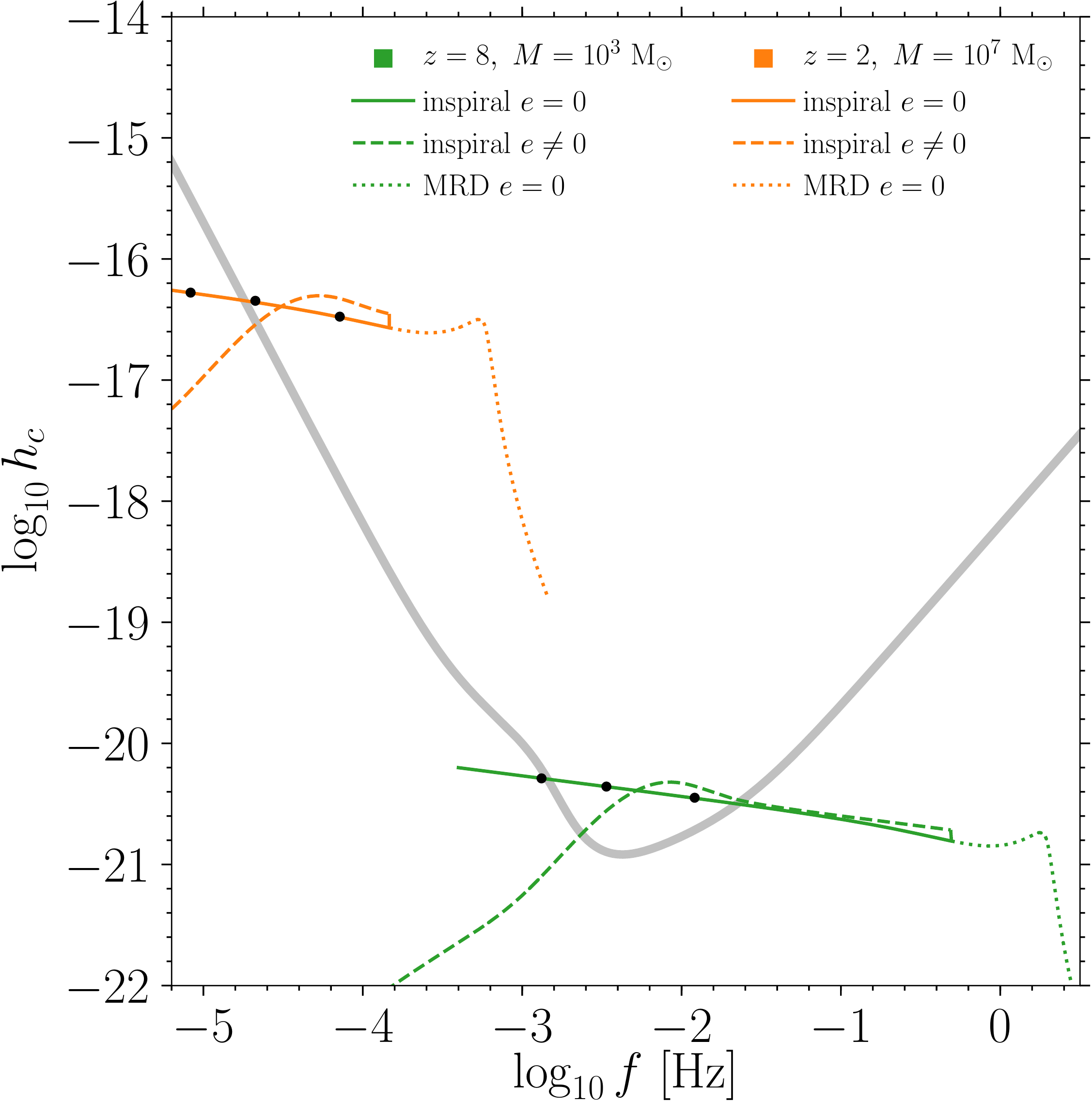}
	\caption{Frequency evolution of the characteristic strains of two equal-mass binaries with total mass $10^3\msun$ (green) and $10^7\msun$ (orange), when eccentricity is considered (dashed lines) or not (solid and dotted lines denote the inspiral and merger-ringdown (MRD) phases respectively). Black dots (from left to right) mark 1 year, 1 month, and 1 day before coalescence. The thick grey line shows the LISA sensitivity curve.} 
	\label{fig:wave_inspiral}
\end{figure}

Finally, in order to evaluate the S/N, we employ the formalism developed by \citet{Barack2004}. The characteristic strain of each harmonic is given by
\begin{equation}
h_{c,n} = \dfrac{1}{\pi d_L} \sqrt{\dfrac{2 G \dot{E_n}}{c^3\dot{f}_n}}\\
\end{equation}
where $\dot{f}_n = n \dot{f}_{\rm orb}$ is the time derivative of the $n$-th harmonic. The total S/N is then computed as 
\begin{equation}\label{eq:SNR}
({\rm S/N})^2 = \sum_n \int \dfrac{h_{c,n}^2}{f_n S_n(f_n)} \ud \ln f_n,
\end{equation}
where $S_n(f_n)$ is the sky-inclination-polarization averaged power spectral density of LISA, which according to its definition in Section~\ref{sec:LISA_sens} already accounts for the fact that the LISA constellation is comprised of two independent interferometers.

\subsubsection{Merger and Ringdown}

The PhenomC model provides a useful approximation to the frequency domain waveforms from circular black hole binaries with non-precessing spins, by ``hybridising'' the inspiral post-Newtonian amplitude and phase with numerical relativity results for the coalescence phase \citep{Santamaria2010}.
However, since they assume circular binaries, these waveforms cannot be used to describe the inspiral phase of the systems in our catalogues, for which eccentricities can be very large in the early inspiral (c.f. Fig.~\ref{fig:e_distr}). However, because the eccentricity
steadily decreases along the MBH inspiral under the effect of GW emission \cite{Peters1964}, all of our sources have an eccentricity $e\lesssim 0.1$ at plunge \citep[c.f. Fig.~\ref{fig:e_isco} and also Fig.~10 of][]{Bonetti2018a}. We can therefore use PhenomC waveforms to approximately compute the S/N of the plunge-merger-ringdown phase and sum this contribution to the inspiral S/N computed as in Section~\ref{sec:snr_insp}. We assume that this phase corresponds to frequencies $>2\,f_{\rm ISCO}$, where $f_{\rm ISCO}$ is the orbital frequency at the last stable circular orbit (ISCO) of a Schwarzschild space-time, i.e. 
\begin{equation}\label{eq:fisco}
f_{\rm ISCO} = \dfrac{1}{2\pi}\dfrac{c^3}{6\sqrt{6} ~G M}.
\end{equation}

\begin{table*}
	\centering
	\caption{Merger rate and population composition of the MBHBs with source-frame chirp mass in the range $10^2-10^8 ~\msun$. Coalescences are grouped according to the evolutionary channel followed, i.e. gas/stellar-driven systems (binaries) or triplet-driven systems (triplets/quadruplets). The numbers in parenthesis refer to the fraction of mergers involving a quadruple system, which for counting purposes, are considered as a sub-class of triplets. Finally, the Triplet-ej entry denotes binaries that started their evolution as part of a triplet, but at the time of coalescence, the third perturbing object has been ejected \citep[see][for a detailed discussion]{Bonetti2018b}. In the following, given the rather small impact of this evolutionary channel on the mass range relevant for LISA, we neglect this sub-sample of sources.}
	\begin{tabular}{c|ccc|cc|ccc|cc}
		\multicolumn{11}{c}{LS}\\
		\hline
		\multirow{3}{*}{$\mathcal{M}$ [$\rm M_{\odot}$]}&\multicolumn{5}{c}{{\it Model-delayed}}&\multicolumn{5}{c}{{\it Model-stalled}}\\
		&\multirow{2}{*}{Rate [yr$^{-1}$]}&\multicolumn{2}{c|}{Binaries}&\multicolumn{2}{c|}{Triplets}&\multirow{2}{*}{Rate [yr$^{-1}$]}&\multicolumn{2}{c|}{Binaries}&\multicolumn{2}{c}{Triplets}\\  
		&& gas &  star & Triplet (quad) & Triplet-ej && gas &  star & Triplet (quad) & Triplet-ej \\
		\hline
		$10^{2}-10^{3}$ & 161.8 & 33.6\% & 65.0\% & 1.4\%(0.03\%) & 0.01\%   & 3.856 & -- & -- & 99.2\%(3.3\%)  & 0.7\% \\
		$10^{3}-10^{4}$ & 49.00 & 54.6\% & 43.4\% & 2.0\%(0.03\%) & 0.1\%    & 10.27 & -- & -- & 97.2\%(11.4\%) & 2.8\% \\
		$10^{4}-10^{5}$ & 8.559 & 47.3\% & 48.0\% & 4.6\%(0.05\%) & 0.1\%    & 5.637 & -- & -- & 95.2\%(27.9\%) & 4.8\% \\
		$10^{5}-10^{6}$ & 3.430 & 25.9\% & 68.4\% & 5.5\%(0.7\%)  & 0.2\%    & 1.436 & -- & -- & 95.5\%(54.8\%) & 4.5\% \\
		$10^{6}-10^{7}$ & 0.468 & 15.3\% & 77.6\% & 6.4\%(0.5\%)  & 0.7\%    & 0.167 & -- & -- & 93.7\%(81.5\%) & 6.3\% \\
		$10^{7}-10^{8}$ & 0.141 & 10.5\% & 80.6\% & 8.3\%(1.0\%)  & 0.7\%    & 0.041 & -- & -- & 92.6\%(83.2\%) & 7.4\% \\
		\hline
		\multicolumn{11}{c}{}\\
		\multicolumn{11}{c}{HS}\\
		\hline
		\multirow{3}{*}{$\mathcal{M}$ [$\rm M_{\odot}$]}&\multicolumn{5}{c}{{\it Model-delayed}}&\multicolumn{5}{c}{{\it Model-stalled}}\\
		&\multirow{2}{*}{Rate [yr$^{-1}$]}&\multicolumn{2}{c|}{Binaries}&\multicolumn{2}{c|}{Triplets}&\multirow{2}{*}{Rate [yr$^{-1}$]}&\multicolumn{2}{c|}{Binaries}&\multicolumn{2}{c}{Triplets}\\  
		&& gas &  star & Triplet (quad) & Triplet-ej && gas &  star & Triplet (quad) & Triplet-ej \\
		\hline
		$10^{2}-10^{3}$ & 0.000 & 0.0 \% & 0.0 \% & 0.0 \%(0.0\%) & 0.0\%    & 0.000 & -- & -- & 0.0  \%(0.0\%)  & 0.0 \% \\
		$10^{3}-10^{4}$ & 0.911 & 14.3\% & 57.7\% & 27.7\%(0.0\%) & 0.0\%    & 0.440 & -- & -- & 100.0\%(0.0\%)  & 0.0 \% \\
		$10^{4}-10^{5}$ & 8.233 & 19.1\% & 34.3\% & 45.6\%(1.5\%) & 1.0\%    & 3.975 & -- & -- & 97.6 \%(3.8\%)  & 2.1 \% \\
		$10^{5}-10^{6}$ & 13.11 & 20.9\% & 39.8\% & 37.4\%(7.4\%) & 1.9\%    & 6.166 & -- & -- & 96.6 \%(22.7\%) & 3.4 \% \\
		$10^{6}-10^{7}$ & 0.610 & 23.2\% & 67.0\% & 9.0 \%(2.2\%) & 0.8\%    & 0.301 & -- & -- & 96.3 \%(57.4\%) & 3.7 \% \\
		$10^{7}-10^{8}$ & 0.100 & 10.7\% & 78.1\% & 10.2\%(1.3\%) & 1.0\%    & 0.056 & -- & -- & 87.3 \%(61.5\%) & 12.7\% \\
		\hline
	\end{tabular}
	\label{tab:tab_rate}
\end{table*}

Examples of the resulting piecewise waveforms in the frequency domain are plotted in Fig.~\ref{fig:wave_inspiral} against the LISA sensitivity. While this procedure is admittedly crude, the waveform is only used to compute indicative source S/N and not for precise parameter estimation. Results are therefore robust against the details of the model. Indeed, we have verified that neglecting the plunge-merger-ringdown S/N does not change the projected LISA detection rates, i.e. typically the additional contribution merely increases
the S/N of sources that would be detected anyway, thus potentially affecting parameter estimation -- especially for the mass and spin of the remnant MBH \citep{Klein2016} -- but not the detection rate. 
The reason for this behaviour lies in the shape of LISA sensitivity, which is maximal for MBHB in the range $10^4-10^6 \msun$ (i.e. the typical mass of LISA events), while is relatively poor for higher masses, where the signal coming from the merger and ringdown is much more relevant than that of the inspiral.

\section{Results}
\label{sec_results}

\subsection{Intrinsic MBHB merger rates}

We start by discussing the intrinsic cosmic MBHB merger rates (i.e. irrespectively from LISA detection capabilities) predicted by the four MBH evolutionary models under investigation. Those are tabulated, divided in source-frame chirp mass decades, in Table~\ref{tab:tab_rate}. We limit the range to $\mathcal{M}<10^8 \msun$ because LISA has poor sensitivity to more massive binaries, and at higher masses the rates drop to $\ll 1$\,yr$^{-1}$ anyway. All results presented in the current section, unless indicated otherwise, refer to models in which we employed the DMM prescription for the gas dynamics, and $\gamma=1$ for the slope of the Denhen's potential. Results obtained with the alternative prescriptions (i.e. HKM and $\gamma = 1.75$) are quantitatively similar.

Differences between the HS and LS models have already been discussed in literature \citep[see, e.g.][]{Sesana2007b,Sesana2011a,Klein2016} and our results are in line with previous findings. Focusing on {\it Model-delayed}, the LS scenario results in much higher merger rates ($\approx 220$\,yr$^{-1}$ vs $\approx 23$\,yr$^{-1}$) with lighter chirp masses in the source frame  (${\cal M}<10^3\msun$ vs ${\cal M}\sim10^5\msun$) compared to the HS case. This is a natural consequence of the seeding mechanism, which features lighter and more abundant seeds in the LS model. The comprehensive dynamical treatment described in Section~\ref{sec:channels} allows us to break down the population in the different channels. The two seeding models are very different also in this respect. In the LS scenario, triplets account only for few \% of the mergers, their occurrence increasing monotonically with mass. Conversely, in the HS scenario, triplets account for about 40\% of the mergers in the mass range most relevant for LISA ($10^4\msun\lesssim {\cal M}\lesssim 10^6\msun$), and have less impact at higher masses. Comparing gas and stellar channels, the latter appears to be more common, reflecting a preference for gas-poor mergers in our models. For both LS and HS, this is particularly true at high masses, where dry mergers account for $\sim 80$\% of the total. At lower masses, the two channels share more similar merger fractions. 

\begin{figure*}
	\includegraphics[scale=0.36,clip=true]{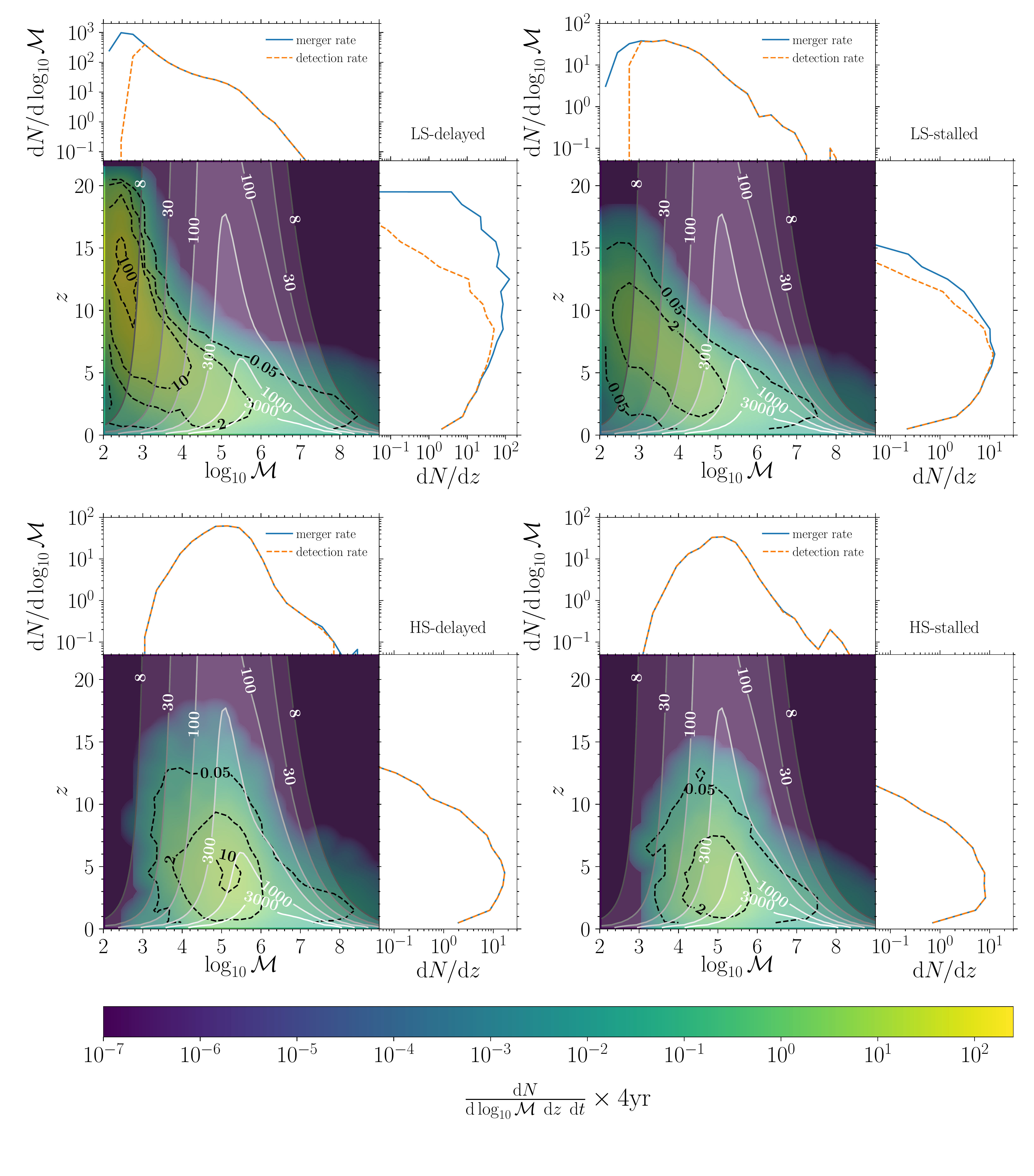}
	\caption{Differential number of mergers (colour gradient and black dashed lines) occurring in the four models as a function of the source-frame chirp mass and redshift during the planned 4-year of LISA lifetime. From the upper left panel, clockwise, we show models LS-{\it delayed}, LS-{\it stalled}, HS-{\it stalled}, and HS-{\it delayed}. Superimposed we show the LISA sensitivity in terms of contours of constant S/N (labelled by white numbers). Note how LISA is essentially blind to low mass, high redshift MBHB coalescences. Finally, for each model, in the upper and right side-panels we show the merger rate (blue line) and detection rate (orange line) distributions marginalised over redshift and chirp mass, respectively. 
	} 
	\label{fig:2Dmergerrate}
\end{figure*}

Comparison between {\it Model-delayed} and {\it Model-stalled} reveals that binary stalling has a much stronger impact on the LS model, for which the merger rate drops by one order of magnitude to $\approx 20$\,yr$^{-1}$. The impact is minor on the HS model, where the rate is only halved, as expected from the fact that a large fraction of mergers was already induced by triplets in the HS {\it Model-delayed}. Also worthy of notice is the statistics of quadruplets. In {\it Model-delayed}, irrespective of the seeding scenario, quadruplets are typically more than ten times rarer than triplets; in {\it Model-stalled}, however, quadruplets can dominate the rate, especially at the high mass end. This is a natural consequence of the underlying physical implementation. The vast majority of quadruplets are indeed formed by two binaries, each coming from a different halo/galaxy. In {\it Model-stalled} a binary can merge only by encountering another systems. As cosmic time proceeds, a progressively larger fraction of MBHs is inevitably in binaries, so that mergers frequently occurs between two MBHBs, resulting in a quadruple system. This occurrence is much rarer in {\it Model-delayed}, where standard merger channels are at work. An important point to make is that even in a pessimistic (from the GW detection perspective) scenario in which all binaries stall, triplets (and quadruplets) provide an efficient channel to keep the expected MBHB cosmic merger rate at a minimum level of few\,yr$^{-1}$. 

\begin{figure*}
	\begin{minipage}{0.47\textwidth}
		\includegraphics[scale=0.38,clip=true]{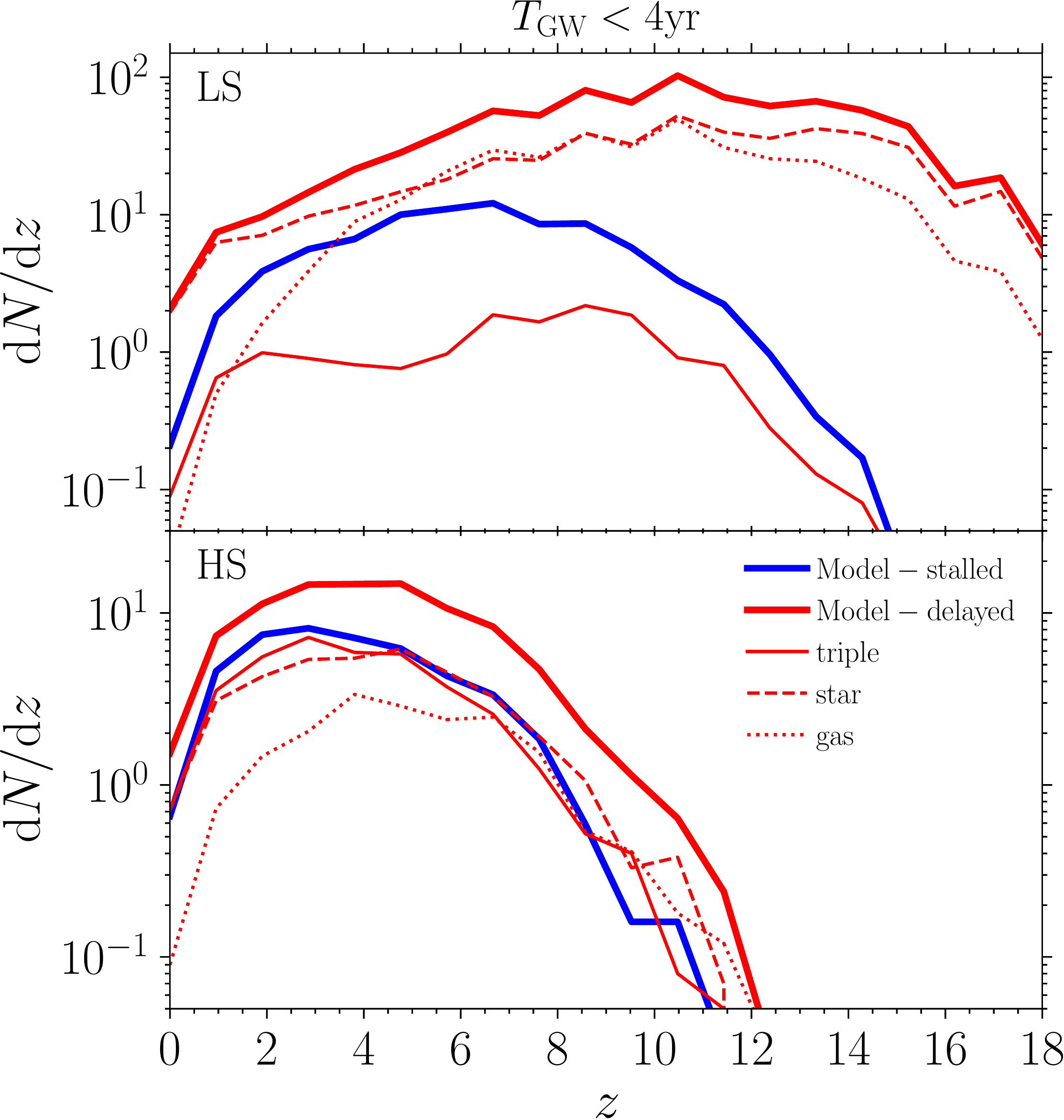}
	\end{minipage}
	\hspace{3mm}
	\begin{minipage}{0.47\textwidth}
		\includegraphics[scale=0.38,clip=true]{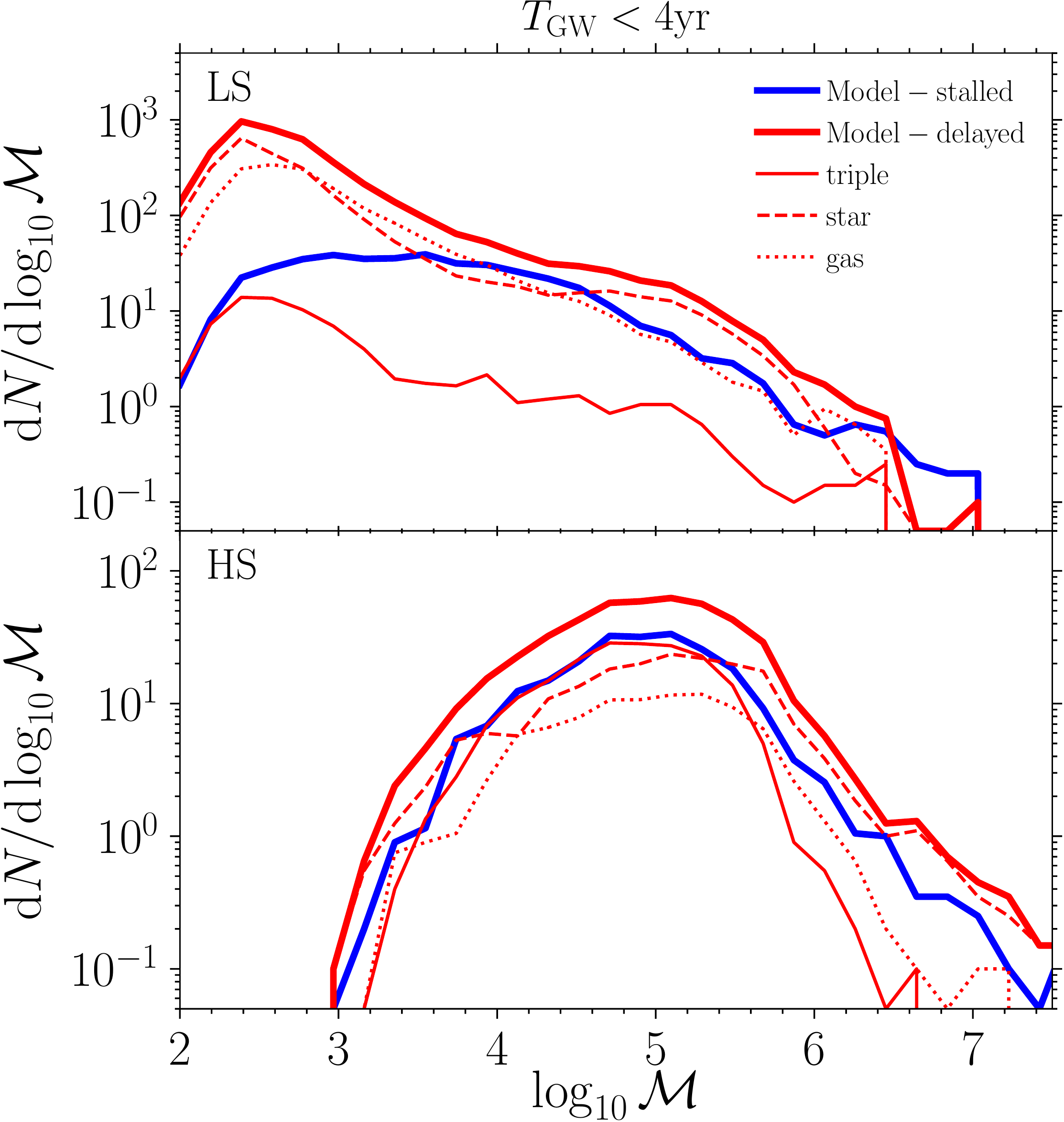}
	\end{minipage}
	\caption{Redshift (left panels) and source-frame chirp mass (right panels) distributions of the differential number of MBHBs coalescing within LISA lifetime, for the LS (upper panels) and HS (lower panels) models. Thick blue lines indicate the events of {\it Model-stalled}, while thick red lines those of {\it Model-delayed}. For {\it Model-delayed}, thin red dashed and dotted lines refer to stellar-driven and gas-driven channels, respectively. The thin red solid line instead represents triplet-induced coalescing MBHBs.}
	\label{fig:merger4yr_distr_Mc_z}
\end{figure*}

Further information can be extracted from Fig.~\ref{fig:2Dmergerrate}, where we report, for all considered models, the differential number of mergers as a function of redshift and source-frame chirp mass, expected in 4 years (the planned LISA mission lifetime). By comparing left to right panels, clearly visible is the reduction and the shift toward lower redshift of the distribution peak. This shift naturally arises from the fact that in {\it Model-stalled} two subsequent galaxy mergers are required to produce a MBHB merger. This effect is more prominent for the LS case (upper panels). The superimposed ``waterfall plot" of S/N contours highlights the window that LISA will open on the cosmic population of MBHBs. Notably, LISA can essentially detect every single MBHB merger in the Universe produced by our HS models (lower panels). In the LS case (upper panels), however, low mass/high redshift systems are beyond LISA capabilities, which is also apparent from the marginalised distributions over masses and redshift, shown respectively by the top and right sub-panel of each triangle plot.

For the sake of clarity, the same marginalised distributions are separately reported in Fig.~\ref{fig:merger4yr_distr_Mc_z}, together with the subdivision into different channels (i.e. star, gas, triplets) for {\it Model-delayed}, where all of them are operational. As already mentioned, in the LS scenario (upper panels) most mergers are stellar-driven and gas-driven (with a slightly predominant role played by the former), while the triplet channel only plays a minor role. The situation is very different in the HS scenario (lower panels), where instead most coalescing MBHBs are either stellar-driven or triple-driven.

\subsection{LISA detection rates}

\begin{figure}
	\includegraphics[scale=0.38,clip=true]{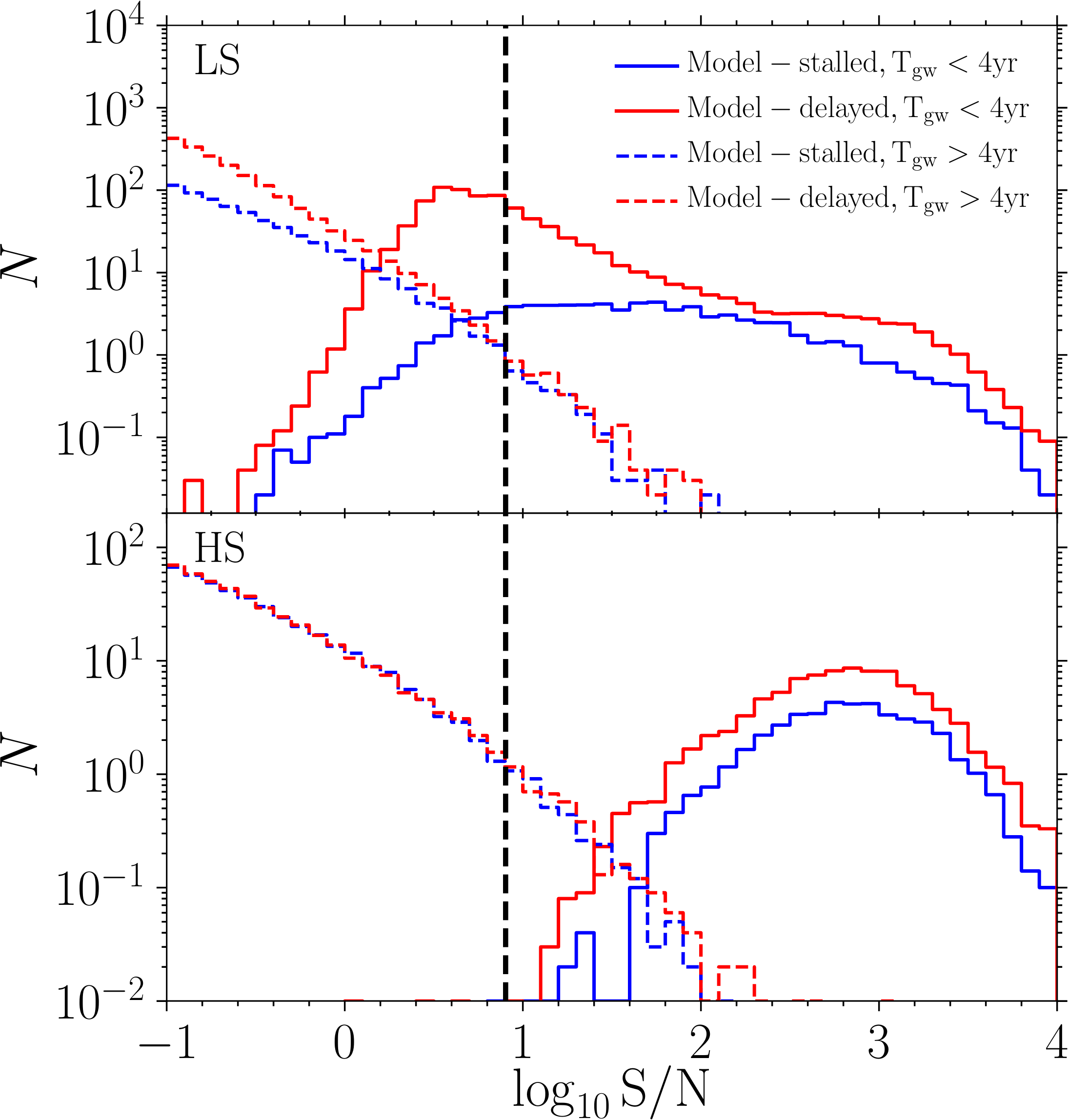}
	\caption{S/N distribution for the LS (upper panel) and HS (lower panel) models. The distribution of MBHBs that merge during the 4 years of LISA lifetime is shown as a solid line, while as a dashed line we show the MBHBs that do not merge within 4 years. The vertical dashed line marks the nominal threshold for LISA detection, $\rm S/N=8$. Colour code as in Fig.~\ref{fig:merger4yr_distr_Mc_z}.}
	\label{fig:SNR_distr}
\end{figure}
\begin{figure*}
	\begin{minipage}{0.47\textwidth}
		\includegraphics[scale=0.38,clip=true]{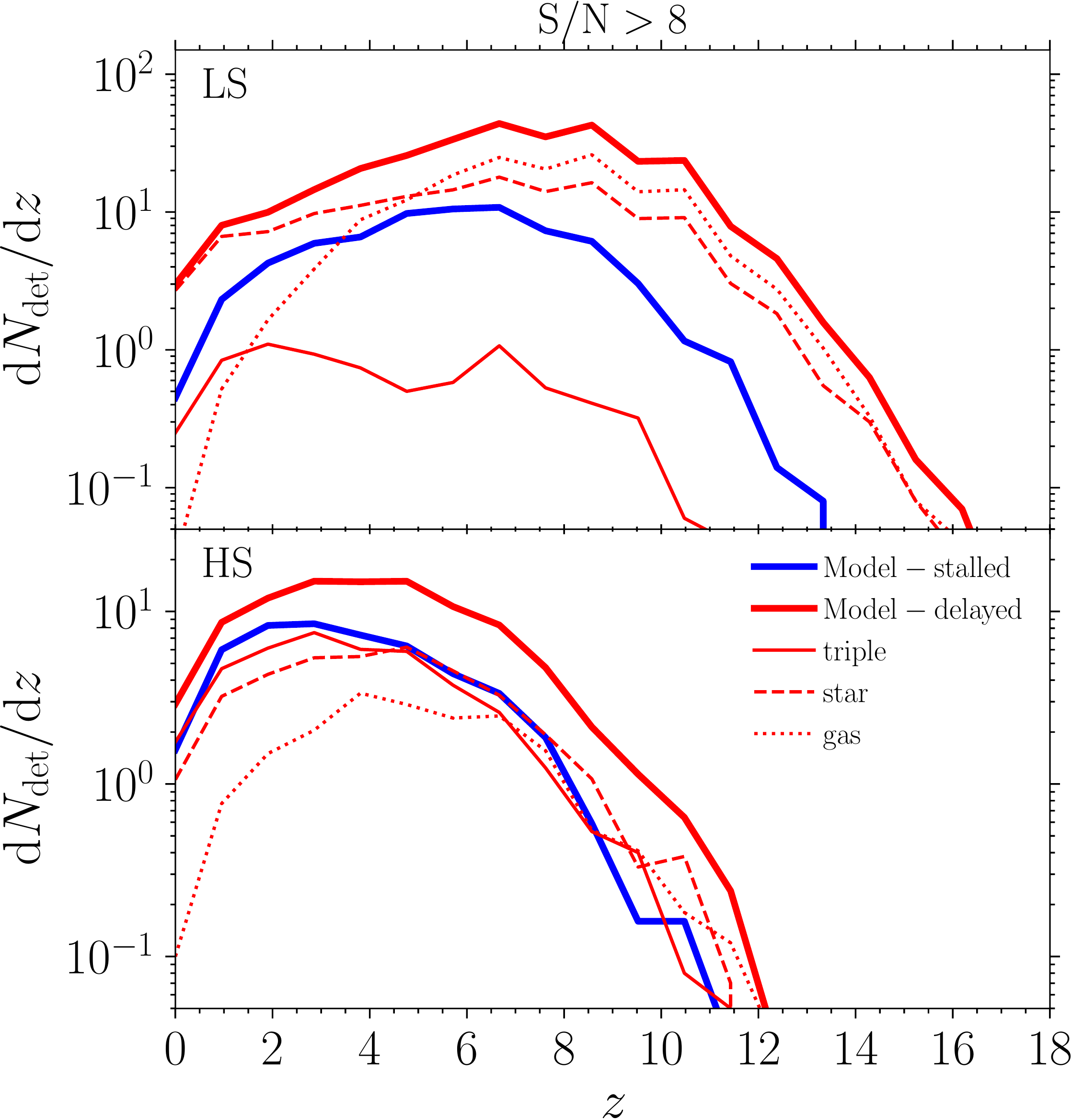}
	\end{minipage}
	\hspace{3mm}
	\begin{minipage}{0.47\textwidth}
		\includegraphics[scale=0.38,clip=true]{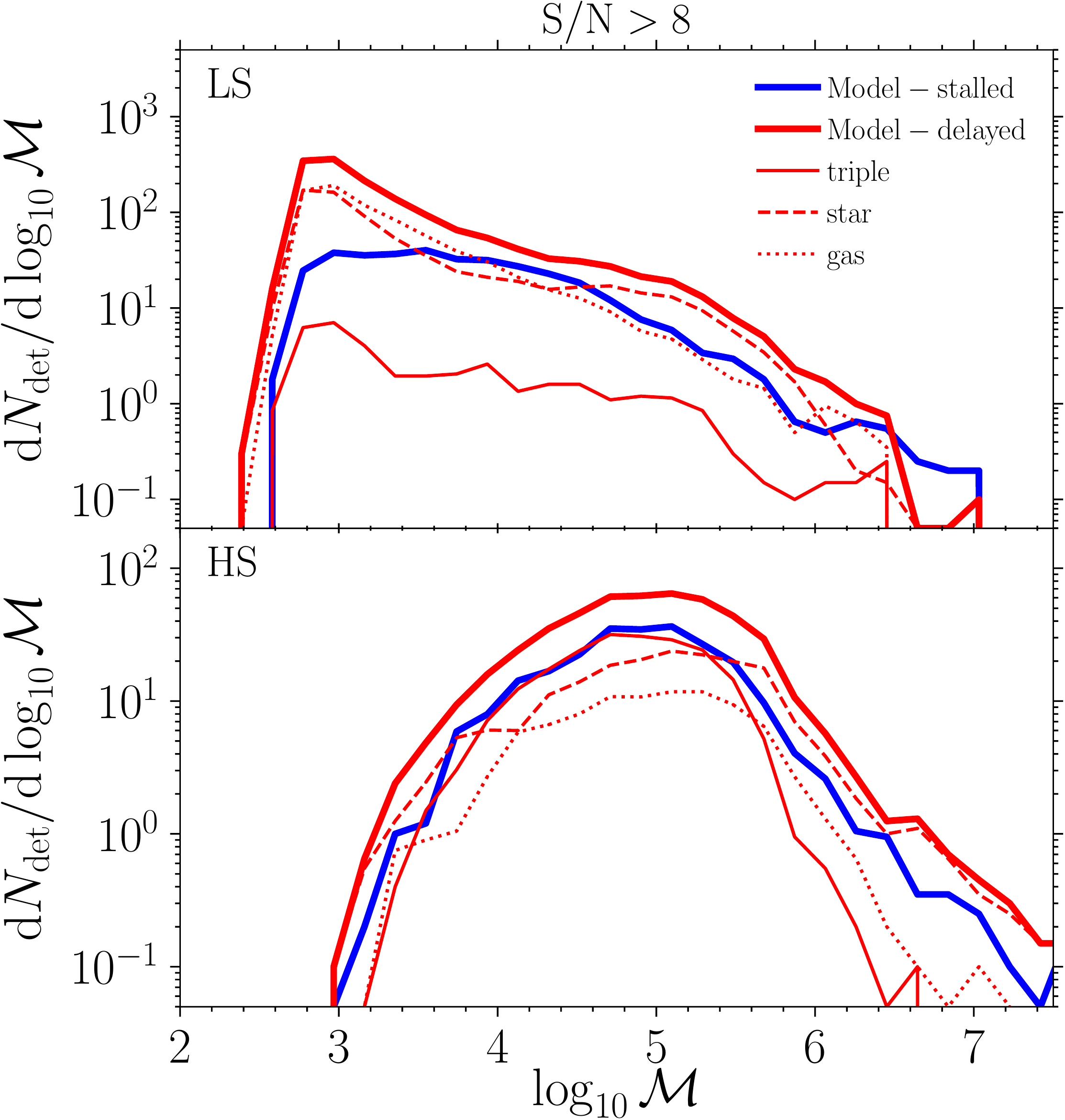}
	\end{minipage}
	\caption{Same as Fig.~\ref{fig:merger4yr_distr_Mc_z}, but considering the distribution of detections exceeding ${\rm S/N}=8$.}
	\label{fig:distr_Mc_z_SNR8}
\end{figure*}

We now focus on the statistics of systems detectable by LISA. Most of them will be a sub-sample of the coalescences discussed in the previous section. However, few detections will involve purely inspiralling binaries. As customary, we consider a nominal detection threshold of ${\rm S/N} = 8$. We evolve the whole sample of MBHBs (equation~\ref{eq:5Ddistribution}) for the 4-year LISA mission lifetime, and we evaluate the S/N of each event employing the formalism described in Section~\ref{sec:GW_emission_detection}. Results are shown in Fig.~\ref{fig:SNR_distr}, where we show all sources with ${\rm S/N} > 0.1$, separating MBHBs that merge within 4 years (solid lines) from the rest (dashed lines). Clearly visible is the difference between the LS (upper panel) and the HS (lower panel) cases. For LS scenarios, in fact, a large fraction of merging binaries does not reach the ${\rm S/N} = 8$ threshold, while for heavy seeds essentially all MBHBs that merge during the mission are detected, as also highlighted by Fig.~\ref{fig:2Dmergerrate} and discussed in the previous section.   

A steady MBHB merger rate has to be supplied by a continuous population of systems evenly distributed in time to coalescence. It is therefore expected that loud binaries merging in the LISA band are only the tip of a much fainter and much larger population of inspiralling systems, far from coalescence. The S/N distribution of this larger population, shown by the dashed lines in Fig.~\ref{fig:SNR_distr}, decays roughly as a power-law, with a tail extending to ${\rm S/N} > 8$. This has a couple of potentially interesting consequences. First, there exists a sub-population of detectable sources with coalescence timescale longer than the nominal LISA lifetime. Although rare (a few at most, in our models), these sources are persistent in the detector data stream for the whole duration of the mission. Second, the large population of inspiralling binaries with ${\rm S/N} < 8$ will produce an unresolved GW background.

\begin{table*}
	\caption{Number of detections ($N_{\rm det}$) from the 3 different evolutionary paths considered. Binaries that merge during the LISA lifetime (left columns) represent the vast majority of events, though we find a few persistent sources that do not coalesce within the 4-year mission (right columns). For stellar and gas-driven MBHBs, comparisons between $\gamma = 1$ and $\gamma = 1.75$ and between DMM and HKM prescriptions (see Section~\ref{sec:channels}) are also reported.}
	\begin{tabular}{l|ccccc|ccccc}
		\hline
		\multirow{2}{*}{Model} & \multicolumn{5}{c|}{$N_{\rm det}(<4 \ \rm yr)$} & \multicolumn{5}{c}{$N_{\rm det}(>4 \ \rm yr)$} \\
		& \multicolumn{2}{c}{star} & \multicolumn{2}{c}{gas} & \multirow{2}{*}{triple} & \multicolumn{2}{c}{star} & \multicolumn{2}{c}{gas} & \multirow{2}{*}{triple} \\
		& $\gamma = 1$ & $\gamma = 1.75$ & DMM & HKM & & $\gamma = 1$ & $\gamma = 1.75$ & DMM & HKM & \\  
		\hline
		LS{\it -stalled} & - &  - & - & - & 67.1 & - &  - & - & - & 2.2 \\
		LS{\it -delayed} & 135.3 & 160.5 & 154.1 & 153.2 & 6.8 & 1.8 & 2.0 & 0.5 & 0.4 & 0.6\\
		HS{\it -stalled} & - &  - & - & - & 44.7 & - &  - & - & - & 3.8 \\
		HS{\it -delayed} & 36.7 & 36.9 & 18.3 & 17.8 & 37.3 & 0.6 & 0.4 & 0.2 & 0.1 & 3.3\\
		\hline
	\end{tabular}
	\label{tab:lisarate}
\end{table*}

The number of LISA detections, assuming a 4-year mission, is listed in Table~\ref{tab:lisarate}, divided into the different evolutionary channels, for all models. We also show the detection rates according to the specific prescriptions adopted for the stellar and gas dynamics (see Section~\ref{sec:channels}). If binary stalling is a common occurrence in Nature (as in {\it Model-stalled}), we predict that LISA will still detect several dozens of merging MBHBs, plus few persistent sources, regardless of the nature of the seeding process. In {\it Model-delayed}, about $\sim 100$ and $\sim 300$ detections are expected in the HS and LS scenarios, respectively. Reflecting the properties of the whole population, in the LS scenario stars and gas contribute evenly to the number of detected systems, with triplets relegated to a mere 3\% of detections. For the HS scenario, detected systems are more evenly distributed among the different channels (although gas is subdominant). Table~\ref{tab:lisarate} shows also how the detection statistics is robust against the different prescriptions considered for stellar and gas-driven dynamics. As already mentioned, we do not find significant differences with the prescription adopted.    

Analogue to Fig.~\ref{fig:merger4yr_distr_Mc_z}, in Fig.~\ref{fig:distr_Mc_z_SNR8} we report, for the LS (upper panels) and HS (lower panels) scenarios, the differential number of detections as a function of redshift (left panel) and source-frame chirp mass (right panel), but considering only detectable sources (i.e. ${\rm S/N} > 8$). For {\it Model-delayed} we show the usual breakup into the three sub-populations of our samples. For the HS model the number of detections essentially coincides with the number of mergers, as LISA, as already discussed, can detect almost all MBHBs in the Universe in this case. Conversely, in the LS scenario, the
LISA detection efficiency is considerably worse, and the number of detections is significantly lower than the total number of mergers. This is due to the low mass, high redshift sources that contribute to the intrinsic merger rate, but which are not loud enough to be observed with LISA.

It is clear that the properties of the detected merging MBHB population can inform us about the MBH seeding mechanism that takes place in Nature \citep[see, e.g.][]{Sesana2011a,Klein2016}. In fact, the production of a large number of mergers of low mass binaries at high redshift is a common signature of LS models, a feature that can easily be distilled from LISA data (see top vs. bottom panels in Fig.~\ref{fig:distr_Mc_z_SNR8}). On the other hand, the impact of a specific dynamical evolution mechanism (e.g. gas vs. stellar-driven) on the merger distribution is more subtle, without clear-cut distinctive features. The challenge of GW astronomy is to reconstruct the MBHB cosmic history from a limited number of detections. Countless ingredients contribute to the outcome \citep[e.g.][]{Berti2008,Sesana2011a,Ricarte2018b}, thus pinning down the correct physics underlying the observed population will require a deep understanding of all the processes at play and a significant reverse-engineering effort. In this context, minor shifts in the redshift distribution predicted by a specific model (see blue vs. red lines in Fig.~\ref{fig:distr_Mc_z_SNR8}) are insufficient for astrophysical inference. Detection of electro-magnetic counterparts will be decisive to assess the importance of the gaseous environment for merging MBHBs. Similarly, distinctive properties of individual merging systems can provide useful information about their nature. We will now see that triplet induced mergers retain a clear signature of their dynamical origin: a non-zero (and possibly very large) eccentricity in the LISA band.

\subsection{Signatures of triplet-induced mergers in LISA data}

\begin{figure}
	\includegraphics[scale=0.38,clip=true]{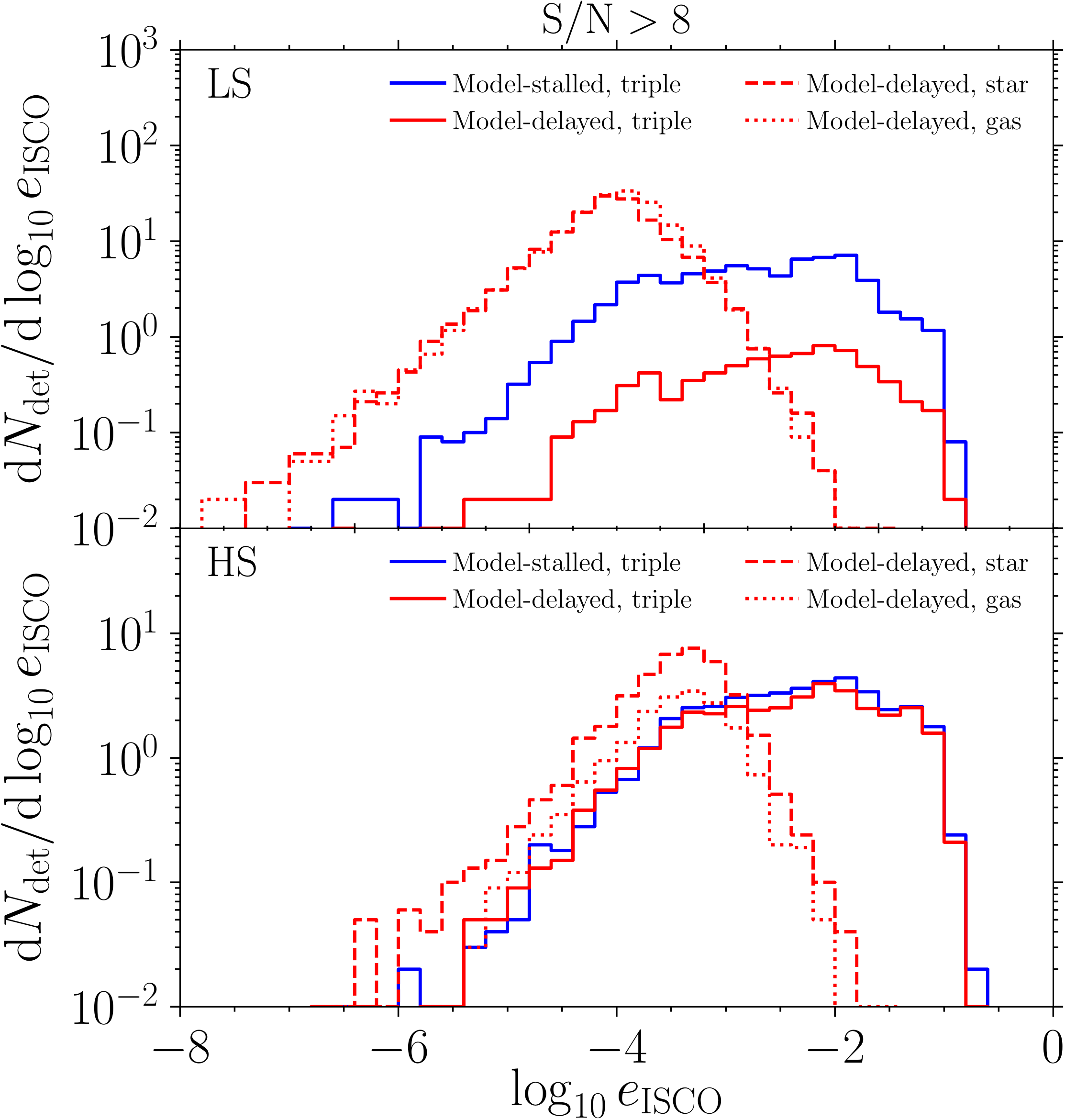}
	\caption{Eccentricity distribution at the ISCO for coalescing MBHBs with ${\rm S/N} > 8$. Colour code and line style as in Fig.~\ref{fig:merger4yr_distr_Mc_z}.}
	\label{fig:e_isco}
\end{figure}

\begin{figure}
	\includegraphics[scale=0.38,clip=true]{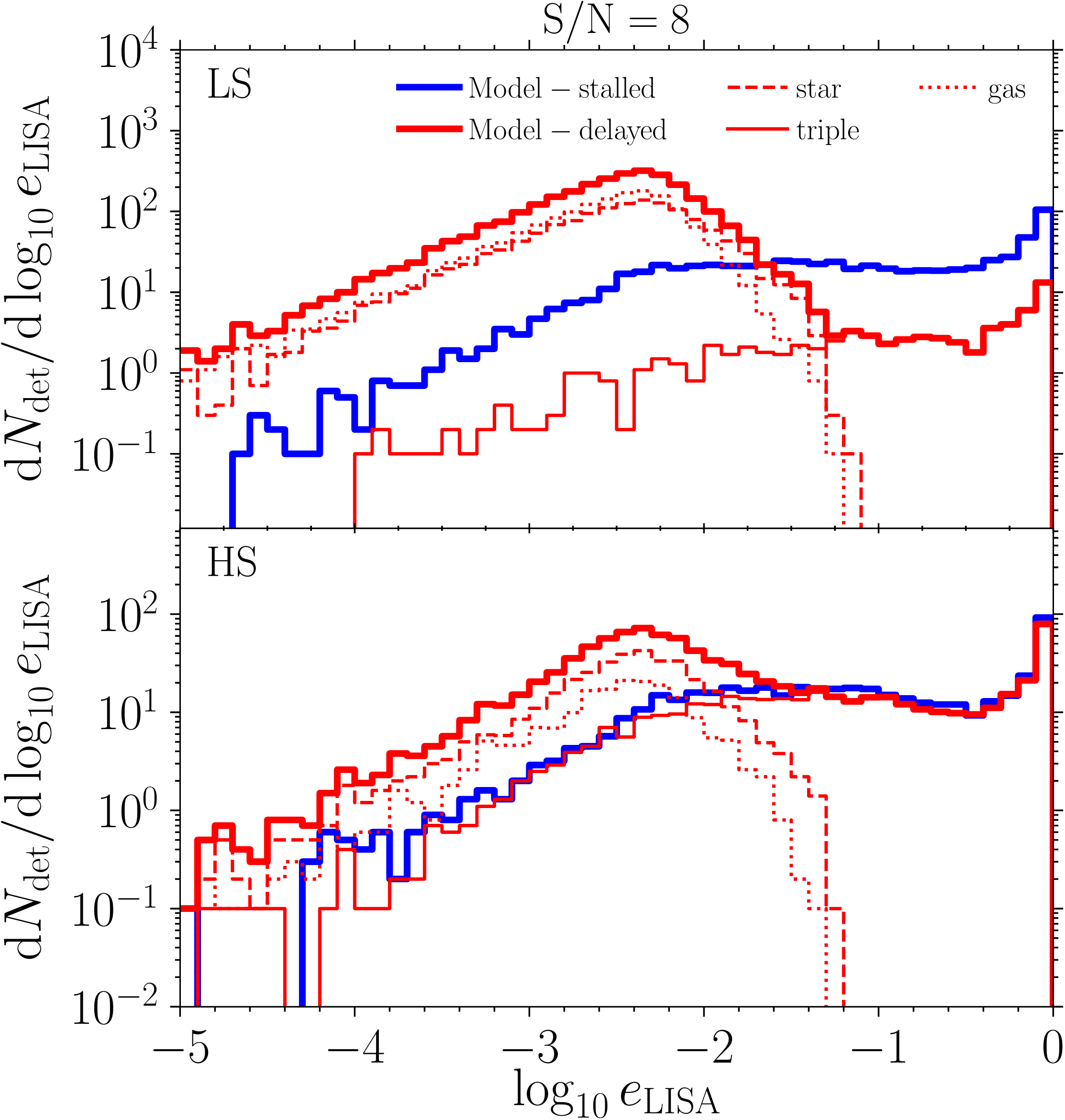}
	\caption{Eccentricity distribution of binaries when the GW signal reaches ${\rm S/N}=8$. The stellar distribution  adopted is the $\gamma=1$ Dehnen's profile, while gas dynamics follows the DMM prescription. Alternative modelling gives quantitatively similar results (see text for details). Colour code and line style as in Fig.~\ref{fig:merger4yr_distr_Mc_z}.}
	\label{fig:e_distr}
\end{figure}
\begin{table}
	\centering
	\caption{Number of detections (in 4 years) of triplet-driven MBHBs with $e_{\rm LISA} > 0.1$ ($N_{\rm det}(e>0.1)$) compared to those belonging to the triplet channel only ($N_{\rm det, triple}$), and to the total number of detections ($N_{\rm det, total}$), for all four considered models.}
	\begin{tabular}{l|ccc}
		\hline\\
		Model	&	$N_{\rm det} (e_{\rm LISA}>0.1)$	&	$N_{\rm det, triple}$&	$N_{\rm det, total}$\\
		\hline
		LS{\it-stalled}	&	31.9	&	69.3	& 69.3   \\
		LS{\it-delayed}	& 	4.2		&	7.4		& 297.7  \\
		HS{\it-stalled}	& 	21.8	& 	48.5	& 48.5   \\
		HS{\it-delayed}	& 	19.3	&	37.3	& 96.4   \\
		\hline
	\end{tabular}
	\label{tab:ecctriple}
\end{table}
\begin{figure}
	\includegraphics[scale=0.305,clip=true]{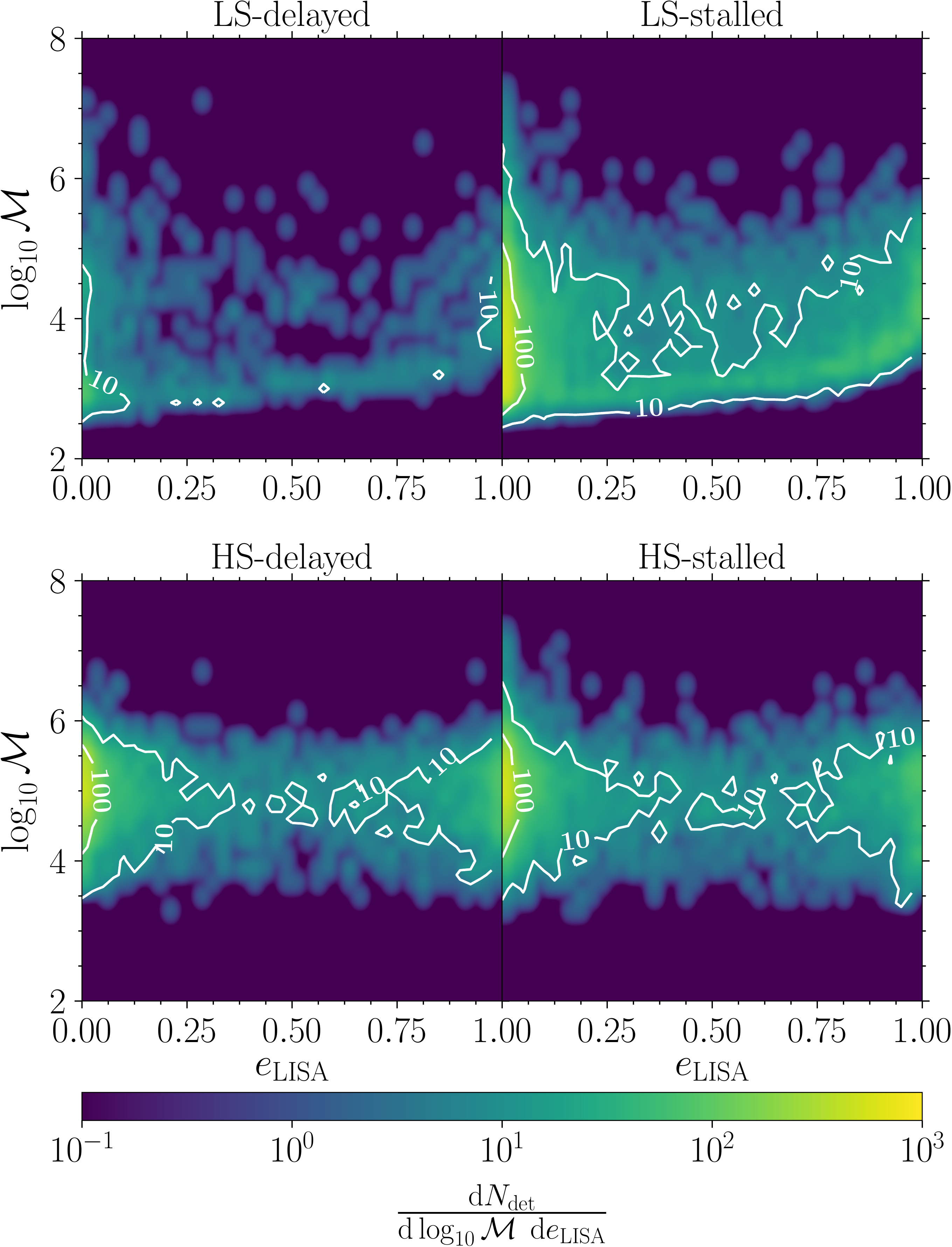}
	\caption{Differential number of merging binaries from the triplet channel and for the four considered models as labelled. We plot the distribution as a function of the eccentricity and source-frame chirp mass when the $\rm S/N=8$ threshold is reached. Note that in the LS case (left panels) the lightest systems are those with the lowest eccentricities.}
	\label{fig:ecc_SNR8_Mc}
\end{figure}

Traditionally, in the study of GW emission from MBHBs and of their observability with LISA, eccentricity has been neglected \citep[e.g.][]{Sesana2004,Klein2016}. The rationale behind this is that GW emission causes efficient orbit circularisation \citep{Peters1964} and unless a binary forms with high eccentricity, any trace of it is washed out by the time the source reaches the LISA band. By modelling the dynamical interplay between MBHBs and their stellar and gaseous environments it was soon realised that binaries can indeed be quite eccentric \citep[e.g.][]{Quinlan1996,Cuadra2009,Khan2012}, and that some residual eccentricity could still be present when a source enters the LISA band \citep{Armitage2005,Sesana2010,Roedig2011,Roedig2012}. As we now discuss, triplets can push eccentricity evolution to the extreme, with dramatic consequences for GW parameter estimation and possibly detection.

We therefore proceed to compare the MBHB eccentricity distribution resulting from the 3 considered evolutionary channels (i.e. star, gas and triplets). The properties of triple-induced mergers are the result of secular dynamics and chaotic interactions that primarily depend on the masses of the involved systems. Conversely, the binary initial conditions and the properties of the surrounding environment play an important role in the standard merger channels, as described in Section~\ref{sec:channels}.  

Since eccentricity changes as the binary shrinks, it is customary to record its value at the last stable orbit, which is a well defined reference. The distribution of $e_{\rm ISCO}$ for all MBHBs that coalesce within 4 years in our models is reported in Fig.~\ref{fig:e_isco}. Here, we assumed model DMM for the gas channel and $\gamma=1$ for the stellar channel. As expected, the final value of the eccentricity right before merger is well below unity in all cases. It is immediately clear, however, that triplets have a distinctive signature, with a relatively flat distribution extending up to $e_{\rm ISCO} \approx 0.1$, more than an order of magnitude larger than what is found in standard stellar- and gas-driven binary evolution.  

Although, in general, MBHBs have $e_{\rm ISCO} \lesssim 0.1$, it should be noted that LISA will observe several thousands of orbital cycles well before final coalescence. The correct identification of a signal early on in the inspiral phase is of critical importance for source pre-localization and electromagnetic follow up planning. It is therefore important to estimate the eccentricity distribution when the signal {\it enters the LISA band}. To this end, during the evaluation of the S/N of each source in our sample, we record the value of the eccentricity at selected S/N thresholds. In particular, we record the properties of the sources when the nominal detection threshold of ${\rm S/N} = 8$ is met. This is defined as the point of entrance of the signal in the LISA band, and the corresponding eccentricity is denoted as $e_{\rm LISA}$. This procedure yields the distributions shown in Fig.~\ref{fig:e_distr}. We compare results from {\it Model-stalled} and {\it Model-delayed}, where for the latter we also visualise the sub-population composition. For both seeding prescriptions (LS in the upper panels and HS in the lower panels), {\it Model-delayed} results in a bimodal eccentricity distribution, with a peak at $\log_{10} e \sim -2.5$, mostly formed by stellar- and gas-driven binaries (cf. thin dashed and dotted red lines) and a plateau stretching up to $e_{\rm LISA} = 1$ (cf. thick blue line and thin solid red line). This latter component is primarily generated by triplet-driven mergers, which can still retain a remarkably high eccentricity when entering the LISA band. The alternative modelling for stellar- and gas-driven binaries (i.e. $\gamma = 1.75$ and HKM prescriptions) provides quantitatively very similar distributions, both in terms of peak position ($\log_{10} e \simeq -2.5$) and cut-off eccentricity ($\log_{10} e \simeq -1$), suggesting that a sizeable residual eccentricity can represent a robust signature of triplet-induced coalescences.

Finally, in Fig.~\ref{fig:ecc_SNR8_Mc} we report, for the triplet channel only, the differential number of events with ${\rm S/N} \geq 8$ as a function of source-frame chirp mass and $e_{\rm LISA}$. The linear scale in eccentricity highlights the bimodality of the distribution, separating those systems that have (low $e_{\rm LISA}$) and those that have not yet (high $e_{\rm LISA}$) undergone circularisation. No significant dependence on the source mass is seen but at ${\cal M}<10^4\msun$, where low eccentricity systems are predominant. This is due to the fact that low mass binaries (especially at high redshift) spend their early inspiral phase in the LISA sensitivity bucket, around 1 mHz. If the eccentricity at this point is too high, the GW power is shifted at higher frequencies, where the LISA sensitivity has already significantly deteriorated, ultimately resulting in a non-detection \citep[see examples in][]{Chen2017}. The statistics of detected eccentric systems is reported in Table~\ref{tab:ecctriple}. Integrating the distribution for $e_{\rm LISA}>0.1$, we expect about 23 (36) and 20 (5) systems for {\it Model-stalled} and {\it Model-delayed}, respectively, assuming the HS (LS) scenario. This implies that, in our models, very eccentric binaries can account for between 1.5\% and 50\% of the total merging MBHBs detectable by LISA.

\section{Discussion and Conclusions}
\label{sec:conclusions}

In this paper we forecasted the expected detection statistics of MBHBs in the context of future space based gravitational wave observatories such as LISA. In particular, we assessed the impact of massive black hole triplets (and quadruplets) on the expected rate of cosmic MBH mergers.

To this end, we employed the machinery developed in the first three papers of the series \citep{Bonetti2016,Bonetti2018a,Bonetti2018b}, which combines a state of the art semi-analytic model for the joint evolution of galaxies and MBHs \citep{Barausse2012} to a large library of dedicated 3-body simulations tailored to study the dynamics of MBH triplets in galactic nuclei \citep{Bonetti2018a}. In our framework, MBHBs have three possible paths to coalescence: hardening against the dense stellar background (stellar channel); efficient extraction of energy and angular momentum by a circumbinary disc (gaseous channel); interaction with a third MBH brought in by a subsequent merger (triplet channel). The efficiency of the stellar and gaseous channels is set by the central stellar density and gas content of the host remnant right after a galaxy merger, which are consistently followed along each branch of the tree. The detailed MBHB dynamical evolution depends, however, on a number of parameters that cannot be determined by the semi-analytic model, such as the $\alpha$ viscosity of the circumbinary disc or the specific slope of the power-law describing the stellar density profile, as well as on the initial binary eccentricity. We thus explored a number of scenarios employing reasonable choices of those parameters.   

We considered two different models for the MBH seeding mechanisms -- low mass seeds as the evolutionary endpoint of Pop III stars (LS), and high mass seeds ($\sim 10^5\msun$) from the collapse of massive protogalactic discs (HS) -- and for each of them we explored two models for the dynamical evolution of the MBHBs -- efficient gas and star driven evolution ({\it Model-delayed}) vs. inefficient gas and star driven evolution  ({\it Model-stalled})-- for a total of 4 main models. In {\it Model-delayed} we considered several prescriptions for the stellar and gaseous channels, as described in Section~\ref{sec:channels}. For each separate channel we constructed the numerical multi-dimensional differential distribution of MBHBs as a function of mass, mass ratio, redshift, (observed) orbital frequency and eccentricity. We then sampled this distribution to obtain Montecarlo realizations of the observable cosmic MBHB population. This procedure allowed us to trace all the GW emitters in the Universe, from loud merging binaries to faint low frequency systems in their early inspiral phase. We used the latest LISA sensitivity curve to study the detectability of this cosmic population and the properties of the detected systems.

For the cosmic MBHB merger and LISA detection rates, our findings are generally consistent with the literature \citep[e.g.][]{Sesana2007b,Sesana2011a,Klein2016}. Considering the fiducial {\it Model-delayed}, Pop III light seeds (LS model) result in high merger rates (hundreds per year) mostly involving low mass systems (${\cal M}<10^{3}\msun$) at high redshift (with a broad peak at $z>10$, extending to $z\approx20$). Conversely, heavy seeds (HS model) produce an order of magnitude fewer mergers ($\approx20$ per year), involving much more massive systems at lower redshift (see Fig.~\ref{fig:2Dmergerrate}). As a consequence, LISA can detect almost all MBHB mergers in the Universe in this latter case, but misses more than 50\% of the systems in the former. The original and more relevant results of this work concern the impact of triplet dynamics on MBHB mergers, GW emission and detection. In this respect, our main findings can be summarised as follows:

\begin{enumerate}
	\item for standard assumptions ({\it Model-delayed}) all three dynamical channels (star, gas, triplets) contribute to the cosmic merger rate of MBHBs. Stars and gas are responsible for a comparable share of mergers both in the LS and in the HS scenarios, whereas the contribution of triplets is more prominent in the HS case (accounting for about 40\% of the total merger rate);  
	\item MBH triplets represent an important ``safety net'' for GW observations at mHz frequencies. {\it Model-stalled} demonstrates that should all binary shrinking mechanisms fail, triple (and quadruple) interactions would still bring a sizeable number of MBHB to final coalescence (about $18\,{\rm yr}^{-1}$ and $10\,{\rm yr}^{-1}$ in models LS and HS respectively); 
	\item even in {\it Model-delayed}, where the fiducial recipes for gas and star driven hardening are employed, triplets are responsible for a non negligible fraction of MBHB coalescences. This is only about 3\% (rate of $\approx2\,{\rm yr}^{-1}$) for the LS model and about 40\% (rate of $\approx10\,{\rm yr}^{-1}$) for the HS model. The different contribution of triplets can be explained by keeping in mind that their formation requires the existence of long-lived MBHBs, and that the greater the binary mass, the longer its lifetime. In the LS scenario, most BH mergers involve low mass objects that still approximately have their original mass. If a higher BH mass is found, then this means that the MBH has experienced several mergers or, most likely, that the additional mass has been accreted. An abundant reservoir of gas usually determines a strong coupling with the environment and, hence, a faster evolution, therefore suppressing the formation of triplets. On the contrary, in the HS scenario, MBHBs start with a higher mass ($\sim 10^5\msun$). This seems to be the sweet spot for triplet-induced mergers, since these binaries are moderately long-lived (as the nuclear gas reservoir is typically negligible compared to these more massive binaries). At the same time, there exists a large population of MBHs with roughly the same mass and thus capable of perturbing the ``stalled'' binaries. This fact also explains why at higher MBH masses the contribution of triplet-induced mergers decreases. Indeed, as the MBHB mass grows, it is less likely that a possible perturber is massive enough to trigger the coalescence;
	\item triplet-induced MBHB inspirals retain remarkably high eccentricities in the LISA band. In all scenarios considered, about 50\% of these systems have an eccentricity $e_{\rm LISA}>0.1$ when entering the LISA band, and eccentricities as large as 0.9 are the norm rather than the exception. Conversely, such high eccentricities are rare among systems evolving via standard shrinking channels (gas and stellar-driven).
\end{enumerate}

Item (ii), in particular, is of paramount importance for future GW observations with LISA. Most of the LISA science case is indeed based on MBHB merger rates derived from semi-analytic models \citep[see e.g. the seminal study of][]{Volonteri2003} with a number of simplifying assumptions, especially concerning the treatment of the MBHB dynamics . On the numerical side, zoom-in high resolution hydrodynamical simulations have been instrumental in understanding the large scale dynamics leading to the pairing of the binaries, but the parsec and sub-parsec evolution still relies on sub grid recipes \citep[e.g,][]{Bellovary2013,Tremmel2018}. Large cosmological simulations, on the other hand, simply do not have yet the resolution to catch the bulk of the expected LISA sources at masses below $10^6\msun$, besides suffering of even more severe shortcomings in the treatment of the MBHB dynamics \citep[see e.g.][and references therein]{Vogelsberger2014,Schaye2015}. Sub-grid recipes are based on analytic modelling and numerical simulations of MBHBs in stellar and gaseous discs. Although the literature on the subject is vast \citep{Berczik2006,Preto2011,Khan2011,Khan2012,Vasiliev2015,Sesana2015,Bortolas2016,Milosavljevic2005,Cuadra2009,Roedig2011,Roedig2012,Shi2012,Farris2014,Tang2018} and the main physical mechanisms at play are generally understood, the evolution outcomes depend on poorly known initial conditions (stellar density, relaxation mechanisms, properties of the gas feeding) and, in the case of gas, on  (micro)physics that is hard to control (e.g. cooling, viscosity, AGN and supernova feedback). Star driven evolution can take up to Gyrs \citep{Vasiliev2015,Sesana2015,Gualandris2017}, whereas the efficiency of coalescence mediated by gas is hard to predict \citep{Lodato2009,Goicovic2018}. A legitimate question is therefore what happens if MBHB hardening mechanisms fail altogether. Our results show that triple MBH interactions are an efficient pathway to bring a sizeable number of MBHBs to coalescence in a cosmological context, thus minimising the risk posed by binary stalling to low frequency GW astronomy.

Item (iv) has important implications both for GW astrophysics and signal detection. The expected occurrence of binaries entering in the LISA band with eccentricities as high as 0.9 calls for the development of faithful eccentric waveforms. For such high eccentricities, detection with circular templates may still be feasible, but will be sub-optimal, causing a significant loss in accumulated S/N. This may ultimately delay detection and may severely affect the estimation of source parameters, jeopardising the chances of success of electromagnetic follow-ups. The development of eccentric waveform is ongoing \citep{Key2011,Nishizawa2016,Huerta2017,Klein2018}, but more work is needed in this direction. In terms of GW astrophysics, the detection of several highly eccentric binaries might be the smoking gun of a substantial population of MBH triplets. In general, standard hardening channels do not produce eccentricities larger than 0.1 in the LISA band \citep{Roedig2012}, although for specific conditions this might be possible. For example, MBHBs hardening in counter rotating stellar systems as well as in counter rotating discs may have extremely high eccentricities \citep{Nixon2011,Sesana2011b,Holley-Bockelmann2015,Schnittman2015}. It is however unclear how relevant these physical configurations are, especially for LISA MBHBs, since the cumulative effect of star-binary interactions/viscous torques tend to align the the orbital MBHB angular momentum to that of the stellar distribution/disk \citep{Gualandris2012,2013ApJ...762...68D,2014MNRAS.439.3476R,2017ApJ...837..135R}.  

Besides MBHBs merging within the LISA mission lifetime, we have also simulated the much larger population of inspiralling systems still far from coalescence. We found that detection of few persistent MBHBs with ${\rm S/N} > 8$ is expected. Moreover, the collective signal of the vast population of sub-threshold systems might give rise to an interesting unresolved background \citep{Sesana2007b}. We plan to study in detail the statistical properties and emitted GW signal of this persistent MBHB population in a forthcoming publication. 

\section*{Acknowledgements}
MB, MC and FH acknowledge partial financial support from the INFN TEONGRAV specific initiative. 
MB acknowledges CINECA, under the TEONGRAV initiative, for the availability of high performance computing resources and support. 
This work was supported by the H2020-MSCA-RISE-2015 Grant No. StronGrHEP-690904. 
This project has received funding (to E. Barausse) from the European
Research Council (ERC) under the European Union’s Horizon 2020
research and innovation programme (grant agreement no. GRAMS-815673;
project title ``GRavity from Astrophysical to Microscopic Scales'').
AS is supported by a University Research Fellowship of the Royal Society. The authors would also like to acknowledge networking support by the COST
Action CA16104.
This work has made use of the Horizon Cluster, hosted by the Institut d'Astrophysique de Paris. 
We thank Stephane Rouberol for running smoothly this cluster for us.

\bibliographystyle{mnras}
\bibliography{biblio} 

\begin{thebibliography}{}
\makeatletter
\relax
\def\mn@urlcharsother{\let\do\@makeother \do\$\do\&\do\#\do\^\do\_\do\%\do\~}
\def\mn@doi{\begingroup\mn@urlcharsother \@ifnextchar [ {\mn@doi@}
  {\mn@doi@[]}}
\def\mn@doi@[#1]#2{\def\@tempa{#1}\ifx\@tempa\@empty \href
  {http://dx.doi.org/#2} {doi:#2}\else \href {http://dx.doi.org/#2} {#1}\fi
  \endgroup}
\def\mn@eprint#1#2{\mn@eprint@#1:#2::\@nil}
\def\mn@eprint@arXiv#1{\href {http://arxiv.org/abs/#1} {{\tt arXiv:#1}}}
\def\mn@eprint@dblp#1{\href {http://dblp.uni-trier.de/rec/bibtex/#1.xml}
  {dblp:#1}}
\def\mn@eprint@#1:#2:#3:#4\@nil{\def\@tempa {#1}\def\@tempb {#2}\def\@tempc
  {#3}\ifx \@tempc \@empty \let \@tempc \@tempb \let \@tempb \@tempa \fi \ifx
  \@tempb \@empty \def\@tempb {arXiv}\fi \@ifundefined
  {mn@eprint@\@tempb}{\@tempb:\@tempc}{\expandafter \expandafter \csname
  mn@eprint@\@tempb\endcsname \expandafter{\@tempc}}}

\bibitem[\protect\citeauthoryear{{Amaro-Seoane}, {Sesana}, {Hoffman},
  {Benacquista}, {Eichhorn}, {Makino}  \& {Spurzem}}{{Amaro-Seoane}
  et~al.}{2010}]{Amaro-Seoane2010}
{Amaro-Seoane} P.,  {Sesana} A.,  {Hoffman} L.,  {Benacquista} M.,  {Eichhorn}
  C.,  {Makino} J.,   {Spurzem} R.,  2010, \mn@doi [\mnras]
  {10.1111/j.1365-2966.2009.16104.x}, \href
  {http://adsabs.harvard.edu/abs/2010MNRAS.402.2308A} {402, 2308}

\bibitem[\protect\citeauthoryear{{Amaro-Seoane} et~al.,}{{Amaro-Seoane}
  et~al.}{2017}]{LISA2017}
{Amaro-Seoane} P.,  et~al., 2017, preprint, \href
  {http://adsabs.harvard.edu/abs/2017arXiv170200786A} {} (\mn@eprint {arXiv}
  {1702.00786})

\bibitem[\protect\citeauthoryear{{Antonini}, {Barausse}  \& {Silk}}{{Antonini}
  et~al.}{2015a}]{Antonini_Barausse2015}
{Antonini} F.,  {Barausse} E.,   {Silk} J.,  2015a, \mn@doi [\apjl]
  {10.1088/2041-8205/806/1/L8}, \href
  {http://adsabs.harvard.edu/abs/2015ApJ...806L...8A} {806, L8}

\bibitem[\protect\citeauthoryear{{Antonini}, {Barausse}  \& {Silk}}{{Antonini}
  et~al.}{2015b}]{Antonini2015}
{Antonini} F.,  {Barausse} E.,   {Silk} J.,  2015b, \mn@doi [\apj]
  {10.1088/0004-637X/812/1/72}, \href
  {http://adsabs.harvard.edu/abs/2015ApJ...812...72A} {812, 72}

\bibitem[\protect\citeauthoryear{{Armitage} \& {Natarajan}}{{Armitage} \&
  {Natarajan}}{2005}]{Armitage2005}
{Armitage} P.~J.,  {Natarajan} P.,  2005, \mn@doi [\apj] {10.1086/497108},
  \href {http://adsabs.harvard.edu/abs/2005ApJ...634..921A} {634, 921}

\bibitem[\protect\citeauthoryear{{Artymowicz} \& {Lubow}}{{Artymowicz} \&
  {Lubow}}{1994}]{Artymowicz1994}
{Artymowicz} P.,  {Lubow} S.~H.,  1994, \mn@doi [\apj] {10.1086/173679}, \href
  {http://adsabs.harvard.edu/abs/1994ApJ...421..651A} {421, 651}

\bibitem[\protect\citeauthoryear{{Ba{\~n}ados} et~al.,}{{Ba{\~n}ados}
  et~al.}{2018}]{Banados2018}
{Ba{\~n}ados} E.,  et~al., 2018, \mn@doi [\nat] {10.1038/nature25180}, \href
  {http://adsabs.harvard.edu/abs/2018Natur.553..473B} {553, 473}

\bibitem[\protect\citeauthoryear{{Babak} et~al.,}{{Babak}
  et~al.}{2017}]{Babak2017}
{Babak} S.,  et~al., 2017, \mn@doi [\prd] {10.1103/PhysRevD.95.103012}, \href
  {http://adsabs.harvard.edu/abs/2017PhRvD..95j3012B} {95, 103012}

\bibitem[\protect\citeauthoryear{{Baes} \& {Dejonghe}}{{Baes} \&
  {Dejonghe}}{2002}]{Baes2002}
{Baes} M.,  {Dejonghe} H.,  2002, \mn@doi [\aap] {10.1051/0004-6361:20021064},
  \href {http://adsabs.harvard.edu/abs/2002A%26A...393..485B} {393, 485}

\bibitem[\protect\citeauthoryear{{Barack} \& {Cutler}}{{Barack} \&
  {Cutler}}{2004}]{Barack2004}
{Barack} L.,  {Cutler} C.,  2004, \mn@doi [\prd] {10.1103/PhysRevD.70.122002},
  \href {http://adsabs.harvard.edu/abs/2004PhRvD..70l2002B} {70, 122002}

\bibitem[\protect\citeauthoryear{{Barausse}}{{Barausse}}{2012}]{Barausse2012}
{Barausse} E.,  2012, \mn@doi [\mnras] {10.1111/j.1365-2966.2012.21057.x},
  \href {http://adsabs.harvard.edu/abs/2012MNRAS.423.2533B} {423, 2533}

\bibitem[\protect\citeauthoryear{{Barausse}, {Shankar}, {Bernardi}, {Dubois}
  \& {Sheth}}{{Barausse} et~al.}{2017}]{Barausse2017}
{Barausse} E.,  {Shankar} F.,  {Bernardi} M.,  {Dubois} Y.,   {Sheth} R.~K.,
  2017, \mn@doi [\mnras] {10.1093/mnras/stx799}, \href
  {http://adsabs.harvard.edu/abs/2017MNRAS.468.4782B} {468, 4782}

\bibitem[\protect\citeauthoryear{{Bellovary}, {Brooks}, {Volonteri},
  {Governato}, {Quinn}  \& {Wadsley}}{{Bellovary} et~al.}{2013}]{Bellovary2013}
{Bellovary} J.,  {Brooks} A.,  {Volonteri} M.,  {Governato} F.,  {Quinn} T.,
  {Wadsley} J.,  2013, \mn@doi [\apj] {10.1088/0004-637X/779/2/136}, \href
  {http://adsabs.harvard.edu/abs/2013ApJ...779..136B} {779, 136}

\bibitem[\protect\citeauthoryear{{Berczik}, {Merritt}, {Spurzem}  \&
  {Bischof}}{{Berczik} et~al.}{2006}]{Berczik2006}
{Berczik} P.,  {Merritt} D.,  {Spurzem} R.,   {Bischof} H.-P.,  2006, \mn@doi
  [\apjl] {10.1086/504426}, \href
  {http://adsabs.harvard.edu/abs/2006ApJ...642L..21B} {642, L21}

\bibitem[\protect\citeauthoryear{{Berti} \& {Volonteri}}{{Berti} \&
  {Volonteri}}{2008}]{Berti2008}
{Berti} E.,  {Volonteri} M.,  2008, \mn@doi [\apj] {10.1086/590379}, \href
  {http://adsabs.harvard.edu/abs/2008ApJ...684..822B} {684, 822}

\bibitem[\protect\citeauthoryear{{Blaes}, {Lee}  \& {Socrates}}{{Blaes}
  et~al.}{2002}]{Blaes2002}
{Blaes} O.,  {Lee} M.~H.,   {Socrates} A.,  2002, \mn@doi [\apj]
  {10.1086/342655}, \href {http://adsabs.harvard.edu/abs/2002ApJ...578..775B}
  {578, 775}

\bibitem[\protect\citeauthoryear{{Bonetti}, {Haardt}, {Sesana}  \&
  {Barausse}}{{Bonetti} et~al.}{2016}]{Bonetti2016}
{Bonetti} M.,  {Haardt} F.,  {Sesana} A.,   {Barausse} E.,  2016, \mn@doi
  [\mnras] {10.1093/mnras/stw1590}, \href
  {http://adsabs.harvard.edu/abs/2016MNRAS.461.4419B} {461, 4419}

\bibitem[\protect\citeauthoryear{{Bonetti}, {Sesana}, {Barausse}  \&
  {Haardt}}{{Bonetti} et~al.}{2018a}]{Bonetti2018b}
{Bonetti} M.,  {Sesana} A.,  {Barausse} E.,   {Haardt} F.,  2018a, \mn@doi
  [\mnras] {10.1093/mnras/sty874}, \href
  {http://adsabs.harvard.edu/abs/2018MNRAS.477.2599B} {477, 2599}

\bibitem[\protect\citeauthoryear{{Bonetti}, {Haardt}, {Sesana}  \&
  {Barausse}}{{Bonetti} et~al.}{2018b}]{Bonetti2018a}
{Bonetti} M.,  {Haardt} F.,  {Sesana} A.,   {Barausse} E.,  2018b, \mn@doi
  [\mnras] {10.1093/mnras/sty896}, \href
  {http://adsabs.harvard.edu/abs/2018MNRAS.477.3910B} {477, 3910}

\bibitem[\protect\citeauthoryear{{Bortolas}, {Gualandris}, {Dotti}, {Spera}  \&
  {Mapelli}}{{Bortolas} et~al.}{2016}]{Bortolas2016}
{Bortolas} E.,  {Gualandris} A.,  {Dotti} M.,  {Spera} M.,   {Mapelli} M.,
  2016, \mn@doi [\mnras] {10.1093/mnras/stw1372}, \href
  {http://adsabs.harvard.edu/abs/2016MNRAS.461.1023B} {461, 1023}

\bibitem[\protect\citeauthoryear{{Boylan-Kolchin}, {Ma}  \&
  {Quataert}}{{Boylan-Kolchin} et~al.}{2008}]{Boylan-Kolchin2008}
{Boylan-Kolchin} M.,  {Ma} C.-P.,   {Quataert} E.,  2008, \mn@doi [\mnras]
  {10.1111/j.1365-2966.2007.12530.x}, \href
  {http://adsabs.harvard.edu/abs/2008MNRAS.383...93B} {383, 93}

\bibitem[\protect\citeauthoryear{{Callegari}, {Mayer}  \&
  {Kazantzidis}}{{Callegari} et~al.}{2008}]{Callegari2008}
{Callegari} S.,  {Mayer} L.,   {Kazantzidis} S.,  2008, \memsai, \href
  {http://cdsads.u-strasbg.fr/abs/2008MmSAI..79.1302C} {79, 1302}

\bibitem[\protect\citeauthoryear{{Callegari}, {Kazantzidis}, {Mayer}, {Colpi},
  {Bellovary}, {Quinn}  \& {Wadsley}}{{Callegari} et~al.}{2011}]{Callegari2011}
{Callegari} S.,  {Kazantzidis} S.,  {Mayer} L.,  {Colpi} M.,  {Bellovary}
  J.~M.,  {Quinn} T.,   {Wadsley} J.,  2011, \mn@doi [\apj]
  {10.1088/0004-637X/729/2/85}, \href
  {http://cdsads.u-strasbg.fr/abs/2011ApJ...729...85C} {729, 85}

\bibitem[\protect\citeauthoryear{{Capelo}, {Volonteri}, {Dotti}, {Bellovary},
  {Mayer}  \& {Governato}}{{Capelo} et~al.}{2015}]{Capelo2015}
{Capelo} P.~R.,  {Volonteri} M.,  {Dotti} M.,  {Bellovary} J.~M.,  {Mayer} L.,
   {Governato} F.,  2015, \mn@doi [\mnras] {10.1093/mnras/stu2500}, \href
  {http://cdsads.u-strasbg.fr/abs/2015MNRAS.447.2123C} {447, 2123}

\bibitem[\protect\citeauthoryear{{Chen} \& {Amaro-Seoane}}{{Chen} \&
  {Amaro-Seoane}}{2017}]{Chen2017}
{Chen} X.,  {Amaro-Seoane} P.,  2017, \mn@doi [\apjl]
  {10.3847/2041-8213/aa74ce}, \href
  {http://adsabs.harvard.edu/abs/2017ApJ...842L...2C} {842, L2}

\bibitem[\protect\citeauthoryear{Colpi}{Colpi}{2014}]{Colpi2014}
Colpi M.,  2014, \mn@doi [Space Sci. Rev.] {10.1007/s11214-014-0067-1}, 183,
  189

\bibitem[\protect\citeauthoryear{{Cornish} \& {Robson}}{{Cornish} \&
  {Robson}}{2018}]{Cornish2018}
{Cornish} N.,  {Robson} T.,  2018, preprint, \href
  {http://adsabs.harvard.edu/abs/2018arXiv180301944C} {} (\mn@eprint {arXiv}
  {1803.01944})

\bibitem[\protect\citeauthoryear{{Cuadra}, {Armitage}, {Alexander}  \&
  {Begelman}}{{Cuadra} et~al.}{2009}]{Cuadra2009}
{Cuadra} J.,  {Armitage} P.~J.,  {Alexander} R.~D.,   {Begelman} M.~C.,  2009,
  \mn@doi [\mnras] {10.1111/j.1365-2966.2008.14147.x}, \href
  {http://adsabs.harvard.edu/abs/2009MNRAS.393.1423C} {393, 1423}

\bibitem[\protect\citeauthoryear{{Dayal}, {Rossi}, {Shiralilou}, {Piana},
  {Choudhury}  \& {Volonteri}}{{Dayal} et~al.}{2019}]{Dayal2019}
{Dayal} P.,  {Rossi} E.~M.,  {Shiralilou} B.,  {Piana} O.,  {Choudhury} T.~R.,
   {Volonteri} M.,  2019, \mn@doi [\mnras] {10.1093/mnras/stz897}, \href
  {http://adsabs.harvard.edu/abs/2019MNRAS.486.2336D} {486, 2336}

\bibitem[\protect\citeauthoryear{{Dehnen}}{{Dehnen}}{1993}]{Dehnen1993}
{Dehnen} W.,  1993, \mn@doi [\mnras] {10.1093/mnras/265.1.250}, \href
  {http://adsabs.harvard.edu/abs/1993MNRAS.265..250D} {265, 250}

\bibitem[\protect\citeauthoryear{{Dosopoulou} \& {Antonini}}{{Dosopoulou} \&
  {Antonini}}{2017}]{Dosopoulou2017}
{Dosopoulou} F.,  {Antonini} F.,  2017, \mn@doi [\apj]
  {10.3847/1538-4357/aa6b58}, \href
  {http://adsabs.harvard.edu/abs/2017ApJ...840...31D} {840, 31}

\bibitem[\protect\citeauthoryear{{Dotti}, {Colpi}, {Haardt}  \&
  {Mayer}}{{Dotti} et~al.}{2007}]{Dotti2007}
{Dotti} M.,  {Colpi} M.,  {Haardt} F.,   {Mayer} L.,  2007, \mn@doi [\mnras]
  {10.1111/j.1365-2966.2007.12010.x}, \href
  {http://adsabs.harvard.edu/abs/2007MNRAS.379..956D} {379, 956}

\bibitem[\protect\citeauthoryear{{Dotti}, {Colpi}, {Pallini}, {Perego}  \&
  {Volonteri}}{{Dotti} et~al.}{2013}]{2013ApJ...762...68D}
{Dotti} M.,  {Colpi} M.,  {Pallini} S.,  {Perego} A.,   {Volonteri} M.,  2013,
  \mn@doi [\apj] {10.1088/0004-637X/762/2/68}, \href
  {http://adsabs.harvard.edu/abs/2013ApJ...762...68D} {762, 68}

\bibitem[\protect\citeauthoryear{{Dotti}, {Merloni}  \& {Montuori}}{{Dotti}
  et~al.}{2015}]{Dotti2015}
{Dotti} M.,  {Merloni} A.,   {Montuori} C.,  2015, \mn@doi [\mnras]
  {10.1093/mnras/stv291}, \href
  {http://adsabs.harvard.edu/abs/2015MNRAS.448.3603D} {448, 3603}

\bibitem[\protect\citeauthoryear{{Dvorkin} \& {Barausse}}{{Dvorkin} \&
  {Barausse}}{2017}]{Dvorkin2017}
{Dvorkin} I.,  {Barausse} E.,  2017, preprint, \href
  {http://adsabs.harvard.edu/abs/2017arXiv170206964D} {} (\mn@eprint {arXiv}
  {1702.06964})

\bibitem[\protect\citeauthoryear{{Enoki}, {Inoue}, {Nagashima}  \&
  {Sugiyama}}{{Enoki} et~al.}{2004}]{Enoki2004}
{Enoki} M.,  {Inoue} K.~T.,  {Nagashima} M.,   {Sugiyama} N.,  2004, \mn@doi
  [\apj] {10.1086/424475}, \href
  {http://adsabs.harvard.edu/abs/2004ApJ...615...19E} {615, 19}

\bibitem[\protect\citeauthoryear{{Farris}, {Duffell}, {MacFadyen}  \&
  {Haiman}}{{Farris} et~al.}{2014}]{Farris2014}
{Farris} B.~D.,  {Duffell} P.,  {MacFadyen} A.~I.,   {Haiman} Z.,  2014,
  \mn@doi [\apj] {10.1088/0004-637X/783/2/134}, \href
  {http://adsabs.harvard.edu/abs/2014ApJ...783..134F} {783, 134}

\bibitem[\protect\citeauthoryear{{Ferrarese} \& {Merritt}}{{Ferrarese} \&
  {Merritt}}{2000}]{Ferrarese2000}
{Ferrarese} L.,  {Merritt} D.,  2000, \mn@doi [\apjl] {10.1086/312838}, \href
  {http://adsabs.harvard.edu/abs/2000ApJ...539L...9F} {539, L9}

\bibitem[\protect\citeauthoryear{{Fiacconi}, {Mayer}, {Ro{\v s}kar}  \&
  {Colpi}}{{Fiacconi} et~al.}{2013}]{Fiacconi2013}
{Fiacconi} D.,  {Mayer} L.,  {Ro{\v s}kar} R.,   {Colpi} M.,  2013, \mn@doi
  [\apjl] {10.1088/2041-8205/777/1/L14}, \href
  {http://adsabs.harvard.edu/abs/2013ApJ...777L..14F} {777, L14}

\bibitem[\protect\citeauthoryear{{Gebhardt} et~al.,}{{Gebhardt}
  et~al.}{2000}]{Gebhardt2000}
{Gebhardt} K.,  et~al., 2000, \mn@doi [\apjl] {10.1086/312840}, \href
  {http://adsabs.harvard.edu/abs/2000ApJ...539L..13G} {539, L13}

\bibitem[\protect\citeauthoryear{{Goicovic}, {Sesana}, {Cuadra}  \&
  {Stasyszyn}}{{Goicovic} et~al.}{2017}]{Goicovic2017}
{Goicovic} F.~G.,  {Sesana} A.,  {Cuadra} J.,   {Stasyszyn} F.,  2017, \mn@doi
  [\mnras] {10.1093/mnras/stx1996}, \href
  {http://adsabs.harvard.edu/abs/2017MNRAS.472..514G} {472, 514}

\bibitem[\protect\citeauthoryear{{Goicovic}, {Maureira-Fredes}, {Sesana},
  {Amaro-Seoane}  \& {Cuadra}}{{Goicovic} et~al.}{2018}]{Goicovic2018}
{Goicovic} F.~G.,  {Maureira-Fredes} C.,  {Sesana} A.,  {Amaro-Seoane} P.,
  {Cuadra} J.,  2018, \mn@doi [\mnras] {10.1093/mnras/sty1709}, \href
  {http://adsabs.harvard.edu/abs/2018MNRAS.479.3438G} {479, 3438}

\bibitem[\protect\citeauthoryear{{Gould} \& {Rix}}{{Gould} \&
  {Rix}}{2000}]{Gould2000}
{Gould} A.,  {Rix} H.-W.,  2000, \mn@doi [\apjl] {10.1086/312562}, \href
  {http://adsabs.harvard.edu/abs/2000ApJ...532L..29G} {532, L29}

\bibitem[\protect\citeauthoryear{{Granato}, {De Zotti}, {Silva}  \&
  {Bressan}}{{Granato} et~al.}{2004}]{Granato2004}
{Granato} G.~L.,  {De Zotti} G.,  {Silva} L.,   {Bressan} A.~and{Danese} L.,
  2004, \mn@doi [\apj] {10.1086/379875}, \href
  {http://adsabs.harvard.edu/abs/2004ApJ...600..580G} {600, 580}

\bibitem[\protect\citeauthoryear{{Gualandris}, {Dotti}  \&
  {Sesana}}{{Gualandris} et~al.}{2012}]{Gualandris2012}
{Gualandris} A.,  {Dotti} M.,   {Sesana} A.,  2012, \mn@doi [\mnras]
  {10.1111/j.1745-3933.2011.01188.x}, \href
  {http://adsabs.harvard.edu/abs/2012MNRAS.420L..38G} {420, L38}

\bibitem[\protect\citeauthoryear{{Gualandris}, {Read}, {Dehnen}  \&
  {Bortolas}}{{Gualandris} et~al.}{2017}]{Gualandris2017}
{Gualandris} A.,  {Read} J.~I.,  {Dehnen} W.,   {Bortolas} E.,  2017, \mn@doi
  [\mnras] {10.1093/mnras/stw2528}, \href
  {http://adsabs.harvard.edu/abs/2017MNRAS.464.2301G} {464, 2301}

\bibitem[\protect\citeauthoryear{{Guillet} \& {Teyssier}}{{Guillet} \&
  {Teyssier}}{2011}]{Guillet2011}
{Guillet} T.,  {Teyssier} R.,  2011, \mn@doi [Journal of Computational Physics]
  {10.1016/j.jcp.2011.02.044}, \href
  {http://adsabs.harvard.edu/abs/2011JCoPh.230.4756G} {230, 4756}

\bibitem[\protect\citeauthoryear{{Haehnelt}}{{Haehnelt}}{1994a}]{Haehnelt1994}
{Haehnelt} M.~G.,  1994a, \mn@doi [\mnras] {10.1093/mnras/269.1.199}, \href
  {http://adsabs.harvard.edu/abs/1994MNRAS.269..199H} {269, 199}

\bibitem[\protect\citeauthoryear{{Haehnelt}}{{Haehnelt}}{1994b}]{1994MNRAS.269..199H}
{Haehnelt} M.~G.,  1994b, \mn@doi [\mnras] {10.1093/mnras/269.1.199}, \href
  {http://adsabs.harvard.edu/abs/1994MNRAS.269..199H} {269, 199}

\bibitem[\protect\citeauthoryear{{Haiman}, {Ciotti}  \& {Ostriker}}{{Haiman}
  et~al.}{2004}]{Haiman2004}
{Haiman} Z.,  {Ciotti} L.,   {Ostriker} J.~P.,  2004, \mn@doi [\apj]
  {10.1086/383022}, \href {http://adsabs.harvard.edu/abs/2004ApJ...606..763H}
  {606, 763}

\bibitem[\protect\citeauthoryear{{Haiman}, {Kocsis}  \& {Menou}}{{Haiman}
  et~al.}{2009}]{Haiman2009}
{Haiman} Z.,  {Kocsis} B.,   {Menou} K.,  2009, \mn@doi [\apj]
  {10.1088/0004-637X/700/2/1952}, \href
  {http://adsabs.harvard.edu/abs/2009ApJ...700.1952H} {700, 1952}

\bibitem[\protect\citeauthoryear{{Heger} \& {Woosley}}{{Heger} \&
  {Woosley}}{2002}]{Heger2002}
{Heger} A.,  {Woosley} S.~E.,  2002, \mn@doi [\apj] {10.1086/338487}, \href
  {http://adsabs.harvard.edu/abs/2002ApJ...567..532H} {567, 532}

\bibitem[\protect\citeauthoryear{{Hein{\"a}m{\"a}ki}}{{Hein{\"a}m{\"a}ki}}{2001}]{Heinamaki2001}
{Hein{\"a}m{\"a}ki} P.,  2001, \mn@doi [\aap] {10.1051/0004-6361:20010460},
  \href {http://adsabs.harvard.edu/abs/2001A%26A...371..795H} {371, 795}

\bibitem[\protect\citeauthoryear{{Hoffman} \& {Loeb}}{{Hoffman} \&
  {Loeb}}{2007}]{Hoffman2007}
{Hoffman} L.,  {Loeb} A.,  2007, \mn@doi [\mnras]
  {10.1111/j.1365-2966.2007.11694.x}, \href
  {http://adsabs.harvard.edu/abs/2007MNRAS.377..957H} {377, 957}

\bibitem[\protect\citeauthoryear{{Holley-Bockelmann} \&
  {Khan}}{{Holley-Bockelmann} \& {Khan}}{2015}]{Holley-Bockelmann2015}
{Holley-Bockelmann} K.,  {Khan} F.~M.,  2015, \mn@doi [\apj]
  {10.1088/0004-637X/810/2/139}, \href
  {http://adsabs.harvard.edu/abs/2015ApJ...810..139H} {810, 139}

\bibitem[\protect\citeauthoryear{{Huerta} et~al.,}{{Huerta}
  et~al.}{2017}]{Huerta2017}
{Huerta} E.~A.,  et~al., 2017, \mn@doi [\prd] {10.1103/PhysRevD.95.024038},
  \href {http://adsabs.harvard.edu/abs/2017PhRvD..95b4038H} {95, 024038}

\bibitem[\protect\citeauthoryear{{Huerta} et~al.,}{{Huerta}
  et~al.}{2018}]{Huerta2018}
{Huerta} E.~A.,  et~al., 2018, \mn@doi [\prd] {10.1103/PhysRevD.97.024031},
  \href {http://adsabs.harvard.edu/abs/2018PhRvD..97b4031H} {97, 024031}

\bibitem[\protect\citeauthoryear{{Jaffe} \& {Backer}}{{Jaffe} \&
  {Backer}}{2003}]{Jaffe2003}
{Jaffe} A.~H.,  {Backer} D.~C.,  2003, \mn@doi [\apj] {10.1086/345443}, \href
  {http://adsabs.harvard.edu/abs/2003ApJ...583..616J} {583, 616}

\bibitem[\protect\citeauthoryear{{Jenet}, {Hobbs}, {Lee}  \&
  {Manchester}}{{Jenet} et~al.}{2005}]{Jenet2005}
{Jenet} F.~A.,  {Hobbs} G.~B.,  {Lee} K.~J.,   {Manchester} R.~N.,  2005,
  \mn@doi [\apjl] {10.1086/431220}, \href
  {http://adsabs.harvard.edu/abs/2005ApJ...625L.123J} {625, L123}

\bibitem[\protect\citeauthoryear{{Kauffmann} \& {Haehnelt}}{{Kauffmann} \&
  {Haehnelt}}{2000}]{Kauffmann2000}
{Kauffmann} G.,  {Haehnelt} M.,  2000, \mn@doi [\mnras]
  {10.1046/j.1365-8711.2000.03077.x}, \href
  {http://adsabs.harvard.edu/abs/2000MNRAS.311..576K} {311, 576}

\bibitem[\protect\citeauthoryear{{Kelley}, {Blecha}, {Hernquist}, {Sesana}  \&
  {Taylor}}{{Kelley} et~al.}{2017}]{Kelley2017}
{Kelley} L.~Z.,  {Blecha} L.,  {Hernquist} L.,  {Sesana} A.,   {Taylor} S.~R.,
  2017, \mn@doi [\mnras] {10.1093/mnras/stx1638}, \href
  {http://adsabs.harvard.edu/abs/2017MNRAS.471.4508K} {471, 4508}

\bibitem[\protect\citeauthoryear{{Key} \& {Cornish}}{{Key} \&
  {Cornish}}{2011}]{Key2011}
{Key} J.~S.,  {Cornish} N.~J.,  2011, \mn@doi [\prd]
  {10.1103/PhysRevD.83.083001}, \href
  {http://adsabs.harvard.edu/abs/2011PhRvD..83h3001K} {83, 083001}

\bibitem[\protect\citeauthoryear{{Khan}, {Just}  \& {Merritt}}{{Khan}
  et~al.}{2011}]{Khan2011}
{Khan} F.~M.,  {Just} A.,   {Merritt} D.,  2011, \mn@doi [\apj]
  {10.1088/0004-637X/732/2/89}, \href
  {http://adsabs.harvard.edu/abs/2011ApJ...732...89K} {732, 89}

\bibitem[\protect\citeauthoryear{{Khan}, {Preto}, {Berczik}, {Berentzen},
  {Just}  \& {Spurzem}}{{Khan} et~al.}{2012a}]{Khan2012}
{Khan} F.~M.,  {Preto} M.,  {Berczik} P.,  {Berentzen} I.,  {Just} A.,
  {Spurzem} R.,  2012a, \mn@doi [\apj] {10.1088/0004-637X/749/2/147}, \href
  {http://adsabs.harvard.edu/abs/2012ApJ...749..147K} {749, 147}

\bibitem[\protect\citeauthoryear{{Khan}, {Berentzen}, {Berczik}, {Just},
  {Mayer}, {Nitadori}  \& {Callegari}}{{Khan} et~al.}{2012b}]{Khan2012b}
{Khan} F.~M.,  {Berentzen} I.,  {Berczik} P.,  {Just} A.,  {Mayer} L.,
  {Nitadori} K.,   {Callegari} S.,  2012b, \mn@doi [\apj]
  {10.1088/0004-637X/756/1/30}, \href
  {http://adsabs.harvard.edu/abs/2012ApJ...756...30K} {756, 30}

\bibitem[\protect\citeauthoryear{{Khan}, {Holley-Bockelmann}, {Berczik}  \&
  {Just}}{{Khan} et~al.}{2013}]{Khan2013}
{Khan} F.~M.,  {Holley-Bockelmann} K.,  {Berczik} P.,   {Just} A.,  2013,
  \mn@doi [\apj] {10.1088/0004-637X/773/2/100}, \href
  {http://cdsads.u-strasbg.fr/abs/2013ApJ...773..100K} {773, 100}

\bibitem[\protect\citeauthoryear{{Khan}, {Fiacconi}, {Mayer}, {Berczik}  \&
  {Just}}{{Khan} et~al.}{2016}]{Khan2016}
{Khan} F.~M.,  {Fiacconi} D.,  {Mayer} L.,  {Berczik} P.,   {Just} A.,  2016,
  \mn@doi [\apj] {10.3847/0004-637X/828/2/73}, \href
  {http://adsabs.harvard.edu/abs/2016ApJ...828...73K} {828, 73}

\bibitem[\protect\citeauthoryear{{Klein} et~al.,}{{Klein}
  et~al.}{2016}]{Klein2016}
{Klein} A.,  et~al., 2016, \mn@doi [\prd] {10.1103/PhysRevD.93.024003}, \href
  {http://adsabs.harvard.edu/abs/2016PhRvD..93b4003K} {93, 024003}

\bibitem[\protect\citeauthoryear{{Klein}, {Boetzel}, {Gopakumar}, {Jetzer}  \&
  {de Vittori}}{{Klein} et~al.}{2018}]{Klein2018}
{Klein} A.,  {Boetzel} Y.,  {Gopakumar} A.,  {Jetzer} P.,   {de Vittori} L.,
  2018, preprint, \href {http://adsabs.harvard.edu/abs/2018arXiv180108542K} {}
  (\mn@eprint {arXiv} {1801.08542})

\bibitem[\protect\citeauthoryear{{Kocsis}, {Haiman}  \& {Loeb}}{{Kocsis}
  et~al.}{2012}]{Kocsis2012}
{Kocsis} B.,  {Haiman} Z.,   {Loeb} A.,  2012, \mn@doi [\mnras]
  {10.1111/j.1365-2966.2012.22118.x}, \href
  {http://adsabs.harvard.edu/abs/2012MNRAS.427.2680K} {427, 2680}

\bibitem[\protect\citeauthoryear{{Kormendy} \& {Ho}}{{Kormendy} \&
  {Ho}}{2013}]{Kormendy2013}
{Kormendy} J.,  {Ho} L.~C.,  2013, \mn@doi [\araa]
  {10.1146/annurev-astro-082708-101811}, \href
  {http://adsabs.harvard.edu/abs/2013ARA%26A..51..511K} {51, 511}

\bibitem[\protect\citeauthoryear{{Kormendy} \& {Richstone}}{{Kormendy} \&
  {Richstone}}{1995}]{Kormendy1995}
{Kormendy} J.,  {Richstone} D.,  1995, \mn@doi [\araa]
  {10.1146/annurev.aa.33.090195.003053}, \href
  {http://adsabs.harvard.edu/abs/1995ARA%26A..33..581K} {33, 581}

\bibitem[\protect\citeauthoryear{{Kozai}}{{Kozai}}{1962}]{Kozai1962}
{Kozai} Y.,  1962, \mn@doi [\aj] {10.1086/108790}, \href
  {http://adsabs.harvard.edu/abs/1962AJ.....67..591K} {67, 591}

\bibitem[\protect\citeauthoryear{{Kulkarni} \& {Loeb}}{{Kulkarni} \&
  {Loeb}}{2012}]{Kulkarni2012}
{Kulkarni} G.,  {Loeb} A.,  2012, \mn@doi [\mnras]
  {10.1111/j.1365-2966.2012.20699.x}, \href
  {http://adsabs.harvard.edu/abs/2012MNRAS.422.1306K} {422, 1306}

\bibitem[\protect\citeauthoryear{{Lapi}, {Raimundo}, {Aversa}, {Cai},
  {Negrello}, {Celotti}, {De Zotti}  \& {Danese}}{{Lapi}
  et~al.}{2014}]{Lapi2014}
{Lapi} A.,  {Raimundo} S.,  {Aversa} R.,  {Cai} Z.-Y.,  {Negrello} M.,
  {Celotti} A.,  {De Zotti} G.,   {Danese} L.,  2014, \mn@doi [\apj]
  {10.1088/0004-637X/782/2/69}, \href
  {http://adsabs.harvard.edu/abs/2014ApJ...782...69L} {782, 69}

\bibitem[\protect\citeauthoryear{{Lidov}}{{Lidov}}{1962}]{Lidov1962}
{Lidov} M.~L.,  1962, \mn@doi [\planss] {10.1016/0032-0633(62)90129-0}, \href
  {http://adsabs.harvard.edu/abs/1962P%26SS....9..719L} {9, 719}

\bibitem[\protect\citeauthoryear{{Lodato}, {Nayakshin}, {King}  \&
  {Pringle}}{{Lodato} et~al.}{2009}]{Lodato2009}
{Lodato} G.,  {Nayakshin} S.,  {King} A.~R.,   {Pringle} J.~E.,  2009, \mn@doi
  [\mnras] {10.1111/j.1365-2966.2009.15179.x}, \href
  {http://adsabs.harvard.edu/abs/2009MNRAS.398.1392L} {398, 1392}

\bibitem[\protect\citeauthoryear{{Lupi}, {Haardt}, {Dotti}  \& {Colpi}}{{Lupi}
  et~al.}{2015}]{Lupi2015}
{Lupi} A.,  {Haardt} F.,  {Dotti} M.,   {Colpi} M.,  2015, \mn@doi [\mnras]
  {10.1093/mnras/stv1920}, \href
  {http://adsabs.harvard.edu/abs/2015MNRAS.453.3437L} {453, 3437}

\bibitem[\protect\citeauthoryear{{Madau} \& {Dickinson}}{{Madau} \&
  {Dickinson}}{2014}]{Madau2014rev}
{Madau} P.,  {Dickinson} M.,  2014, \mn@doi [\araa]
  {10.1146/annurev-astro-081811-125615}, \href
  {http://adsabs.harvard.edu/abs/2014ARA%26A..52..415M} {52, 415}

\bibitem[\protect\citeauthoryear{{Madau} \& {Rees}}{{Madau} \&
  {Rees}}{2001}]{Madau2001}
{Madau} P.,  {Rees} M.~J.,  2001, \mn@doi [\apjl] {10.1086/319848}, \href
  {http://adsabs.harvard.edu/abs/2001ApJ...551L..27M} {551, L27}

\bibitem[\protect\citeauthoryear{{Magorrian} et~al.,}{{Magorrian}
  et~al.}{1998}]{Magorrian1998}
{Magorrian} J.,  et~al., 1998, \mn@doi [\aj] {10.1086/300353}, \href
  {http://adsabs.harvard.edu/abs/1998AJ....115.2285M} {115, 2285}

\bibitem[\protect\citeauthoryear{{Mayer}, {Kazantzidis}, {Madau}, {Colpi},
  {Quinn}  \& {Wadsley}}{{Mayer} et~al.}{2007}]{Mayer2007}
{Mayer} L.,  {Kazantzidis} S.,  {Madau} P.,  {Colpi} M.,  {Quinn} T.,
  {Wadsley} J.,  2007, \mn@doi [Science] {10.1126/science.1141858}, \href
  {http://cdsads.u-strasbg.fr/abs/2007Sci...316.1874M} {316, 1874}

\bibitem[\protect\citeauthoryear{{Mayer}, {Tamburello}, {Lupi}, {Keller},
  {Wadsley}  \& {Madau}}{{Mayer} et~al.}{2016}]{Mayer2016}
{Mayer} L.,  {Tamburello} V.,  {Lupi} A.,  {Keller} B.,  {Wadsley} J.,
  {Madau} P.,  2016, \mn@doi [\apjl] {10.3847/2041-8205/830/1/L13}, \href
  {http://cdsads.u-strasbg.fr/abs/2016ApJ...830L..13M} {830, L13}

\bibitem[\protect\citeauthoryear{{Mikkola} \& {Valtonen}}{{Mikkola} \&
  {Valtonen}}{1990}]{Mikkola1990}
{Mikkola} S.,  {Valtonen} M.~J.,  1990, \mn@doi [\apj] {10.1086/168250}, \href
  {http://adsabs.harvard.edu/abs/1990ApJ...348..412M} {348, 412}

\bibitem[\protect\citeauthoryear{{Mikkola} \& {Valtonen}}{{Mikkola} \&
  {Valtonen}}{1992}]{Mikkola1992}
{Mikkola} S.,  {Valtonen} M.~J.,  1992, \mn@doi [\mnras]
  {10.1093/mnras/259.1.115}, \href
  {http://adsabs.harvard.edu/abs/1992MNRAS.259..115M} {259, 115}

\bibitem[\protect\citeauthoryear{{Milosavljevi{\'c}} \&
  {Phinney}}{{Milosavljevi{\'c}} \& {Phinney}}{2005}]{Milosavljevic2005}
{Milosavljevi{\'c}} M.,  {Phinney} E.~S.,  2005, \mn@doi [\apjl]
  {10.1086/429618}, \href {http://adsabs.harvard.edu/abs/2005ApJ...622L..93M}
  {622, L93}

\bibitem[\protect\citeauthoryear{{Mortlock} et~al.,}{{Mortlock}
  et~al.}{2011}]{Mortlock2011}
{Mortlock} D.~J.,  et~al., 2011, \mn@doi [\nat] {10.1038/nature10159}, \href
  {http://adsabs.harvard.edu/abs/2011Natur.474..616M} {474, 616}

\bibitem[\protect\citeauthoryear{{Nishizawa}, {Berti}, {Klein}  \&
  {Sesana}}{{Nishizawa} et~al.}{2016}]{Nishizawa2016}
{Nishizawa} A.,  {Berti} E.,  {Klein} A.,   {Sesana} A.,  2016, \mn@doi [\prd]
  {10.1103/PhysRevD.94.064020}, \href
  {http://adsabs.harvard.edu/abs/2016PhRvD..94f4020N} {94, 064020}

\bibitem[\protect\citeauthoryear{{Nixon}, {Cossins}, {King}  \&
  {Pringle}}{{Nixon} et~al.}{2011}]{Nixon2011}
{Nixon} C.~J.,  {Cossins} P.~J.,  {King} A.~R.,   {Pringle} J.~E.,  2011,
  \mn@doi [\mnras] {10.1111/j.1365-2966.2010.17952.x}, \href
  {http://adsabs.harvard.edu/abs/2011MNRAS.412.1591N} {412, 1591}

\bibitem[\protect\citeauthoryear{{Parkinson}, {Cole}  \& {Helly}}{{Parkinson}
  et~al.}{2008}]{Parkinson2008}
{Parkinson} H.,  {Cole} S.,   {Helly} J.,  2008, \mn@doi [\mnras]
  {10.1111/j.1365-2966.2007.12517.x}, \href
  {http://adsabs.harvard.edu/abs/2008MNRAS.383..557P} {383, 557}

\bibitem[\protect\citeauthoryear{Peters}{Peters}{1964}]{Peters1964}
Peters P.~C.,  1964, \mn@doi [Phys. Rev.] {10.1103/PhysRev.136.B1224}, 136,
  B1224

\bibitem[\protect\citeauthoryear{Peters \& Mathews}{Peters \&
  Mathews}{1963}]{Peters1963}
Peters P.~C.,  Mathews J.,  1963, \mn@doi [Phys. Rev.]
  {10.1103/PhysRev.131.435}, 131, 435

\bibitem[\protect\citeauthoryear{{Press} \& {Schechter}}{{Press} \&
  {Schechter}}{1974}]{Press1974}
{Press} W.~H.,  {Schechter} P.,  1974, \mn@doi [\apj] {10.1086/152650}, \href
  {http://adsabs.harvard.edu/abs/1974ApJ...187..425P} {187, 425}

\bibitem[\protect\citeauthoryear{{Preto}, {Berentzen}, {Berczik}  \&
  {Spurzem}}{{Preto} et~al.}{2011}]{Preto2011}
{Preto} M.,  {Berentzen} I.,  {Berczik} P.,   {Spurzem} R.,  2011, \mn@doi
  [\apjl] {10.1088/2041-8205/732/2/L26}, \href
  {http://adsabs.harvard.edu/abs/2011ApJ...732L..26P} {732, L26}

\bibitem[\protect\citeauthoryear{{Quinlan}}{{Quinlan}}{1996}]{Quinlan1996}
{Quinlan} G.~D.,  1996, \mn@doi [\na] {10.1016/S1384-1076(96)00003-6}, \href
  {http://adsabs.harvard.edu/abs/1996NewA....1...35Q} {1, 35}

\bibitem[\protect\citeauthoryear{{Rasskazov} \& {Merritt}}{{Rasskazov} \&
  {Merritt}}{2017}]{2017ApJ...837..135R}
{Rasskazov} A.,  {Merritt} D.,  2017, \mn@doi [\apj]
  {10.3847/1538-4357/aa6188}, \href
  {http://adsabs.harvard.edu/abs/2017ApJ...837..135R} {837, 135}

\bibitem[\protect\citeauthoryear{{Rhook} \& {Wyithe}}{{Rhook} \&
  {Wyithe}}{2005}]{Rhook2005}
{Rhook} K.~J.,  {Wyithe} J.~S.~B.,  2005, \mn@doi [\mnras]
  {10.1111/j.1365-2966.2005.08987.x}, \href
  {http://adsabs.harvard.edu/abs/2005MNRAS.361.1145R} {361, 1145}

\bibitem[\protect\citeauthoryear{{Ricarte} \& {Natarajan}}{{Ricarte} \&
  {Natarajan}}{2018a}]{Ricarte2018a}
{Ricarte} A.,  {Natarajan} P.,  2018a, \mn@doi [\mnras]
  {10.1093/mnras/stx2851}, \href
  {http://adsabs.harvard.edu/abs/2018MNRAS.474.1995R} {474, 1995}

\bibitem[\protect\citeauthoryear{{Ricarte} \& {Natarajan}}{{Ricarte} \&
  {Natarajan}}{2018b}]{Ricarte2018b}
{Ricarte} A.,  {Natarajan} P.,  2018b, \mn@doi [\mnras]
  {10.1093/mnras/sty2448}, \href
  {http://adsabs.harvard.edu/abs/2018MNRAS.481.3278R} {481, 3278}

\bibitem[\protect\citeauthoryear{{Rodriguez-Gomez} et~al.,}{{Rodriguez-Gomez}
  et~al.}{2015}]{Rodriguez-Gomez2015}
{Rodriguez-Gomez} V.,  et~al., 2015, \mn@doi [\mnras] {10.1093/mnras/stv264},
  \href {http://adsabs.harvard.edu/abs/2015MNRAS.449...49R} {449, 49}

\bibitem[\protect\citeauthoryear{{Roedig} \& {Sesana}}{{Roedig} \&
  {Sesana}}{2012}]{Roedig2012}
{Roedig} C.,  {Sesana} A.,  2012, in Journal of Physics Conference Series. p.
  012035 (\mn@eprint {arXiv} {1111.3742}),
  \mn@doi{10.1088/1742-6596/363/1/012035}

\bibitem[\protect\citeauthoryear{{Roedig} \& {Sesana}}{{Roedig} \&
  {Sesana}}{2014}]{2014MNRAS.439.3476R}
{Roedig} C.,  {Sesana} A.,  2014, \mn@doi [\mnras] {10.1093/mnras/stu194},
  \href {http://adsabs.harvard.edu/abs/2014MNRAS.439.3476R} {439, 3476}

\bibitem[\protect\citeauthoryear{{Roedig}, {Dotti}, {Sesana}, {Cuadra}  \&
  {Colpi}}{{Roedig} et~al.}{2011}]{Roedig2011}
{Roedig} C.,  {Dotti} M.,  {Sesana} A.,  {Cuadra} J.,   {Colpi} M.,  2011,
  \mn@doi [\mnras] {10.1111/j.1365-2966.2011.18927.x}, \href
  {http://adsabs.harvard.edu/abs/2011MNRAS.415.3033R} {415, 3033}

\bibitem[\protect\citeauthoryear{{Ro{\v s}kar}, {Fiacconi}, {Mayer},
  {Kazantzidis}, {Quinn}  \& {Wadsley}}{{Ro{\v s}kar}
  et~al.}{2015}]{Roskar2015}
{Ro{\v s}kar} R.,  {Fiacconi} D.,  {Mayer} L.,  {Kazantzidis} S.,  {Quinn}
  T.~R.,   {Wadsley} J.,  2015, \mn@doi [\mnras] {10.1093/mnras/stv312}, \href
  {http://adsabs.harvard.edu/abs/2015MNRAS.449..494R} {449, 494}

\bibitem[\protect\citeauthoryear{{Ryu}, {Perna}, {Haiman}, {Ostriker}  \&
  {Stone}}{{Ryu} et~al.}{2018}]{Ryu2017}
{Ryu} T.,  {Perna} R.,  {Haiman} Z.,  {Ostriker} J.~P.,   {Stone} N.~C.,  2018,
  \mn@doi [\mnras] {10.1093/mnras/stx2524}, \href
  {http://adsabs.harvard.edu/abs/2018MNRAS.473.3410R} {473, 3410}

\bibitem[\protect\citeauthoryear{{Santamar{\'{\i}}a}
  et~al.,}{{Santamar{\'{\i}}a} et~al.}{2010}]{Santamaria2010}
{Santamar{\'{\i}}a} L.,  et~al., 2010, \mn@doi [\prd]
  {10.1103/PhysRevD.82.064016}, \href
  {http://adsabs.harvard.edu/abs/2010PhRvD..82f4016S} {82, 064016}

\bibitem[\protect\citeauthoryear{{Schaye} et~al.,}{{Schaye}
  et~al.}{2015}]{Schaye2015}
{Schaye} J.,  et~al., 2015, \mn@doi [\mnras] {10.1093/mnras/stu2058}, \href
  {http://adsabs.harvard.edu/abs/2015MNRAS.446..521S} {446, 521}

\bibitem[\protect\citeauthoryear{{Schnittman} \& {Krolik}}{{Schnittman} \&
  {Krolik}}{2015}]{Schnittman2015}
{Schnittman} J.~D.,  {Krolik} J.~H.,  2015, \mn@doi [\apj]
  {10.1088/0004-637X/806/1/88}, \href
  {http://adsabs.harvard.edu/abs/2015ApJ...806...88S} {806, 88}

\bibitem[\protect\citeauthoryear{{Sesana}}{{Sesana}}{2010}]{Sesana2010}
{Sesana} A.,  2010, \mn@doi [\apj] {10.1088/0004-637X/719/1/851}, \href
  {http://adsabs.harvard.edu/abs/2010ApJ...719..851S} {719, 851}

\bibitem[\protect\citeauthoryear{{Sesana} \& {Khan}}{{Sesana} \&
  {Khan}}{2015}]{Sesana2015}
{Sesana} A.,  {Khan} F.~M.,  2015, \mn@doi [\mnras] {10.1093/mnrasl/slv131},
  \href {http://adsabs.harvard.edu/abs/2015MNRAS.454L..66S} {454, L66}

\bibitem[\protect\citeauthoryear{{Sesana}, {Haardt}, {Madau}  \&
  {Volonteri}}{{Sesana} et~al.}{2004}]{Sesana2004}
{Sesana} A.,  {Haardt} F.,  {Madau} P.,   {Volonteri} M.,  2004, \mn@doi [\apj]
  {10.1086/422185}, \href {http://adsabs.harvard.edu/abs/2004ApJ...611..623S}
  {611, 623}

\bibitem[\protect\citeauthoryear{{Sesana}, {Haardt}, {Madau}  \&
  {Volonteri}}{{Sesana} et~al.}{2005}]{Sesana2005}
{Sesana} A.,  {Haardt} F.,  {Madau} P.,   {Volonteri} M.,  2005, \mn@doi [\apj]
  {10.1086/428492}, \href {http://adsabs.harvard.edu/abs/2005ApJ...623...23S}
  {623, 23}

\bibitem[\protect\citeauthoryear{{Sesana}, {Haardt}  \& {Madau}}{{Sesana}
  et~al.}{2006}]{Sesana2006}
{Sesana} A.,  {Haardt} F.,   {Madau} P.,  2006, \mn@doi [\apj]
  {10.1086/507596}, \href {http://adsabs.harvard.edu/abs/2006ApJ...651..392S}
  {651, 392}

\bibitem[\protect\citeauthoryear{{Sesana}, {Volonteri}  \& {Haardt}}{{Sesana}
  et~al.}{2007}]{Sesana2007b}
{Sesana} A.,  {Volonteri} M.,   {Haardt} F.,  2007, \mn@doi [\mnras]
  {10.1111/j.1365-2966.2007.11734.x}, \href
  {http://adsabs.harvard.edu/abs/2007MNRAS.377.1711S} {377, 1711}

\bibitem[\protect\citeauthoryear{{Sesana}, {Vecchio}  \& {Colacino}}{{Sesana}
  et~al.}{2008}]{Sesana_Vecchio2008}
{Sesana} A.,  {Vecchio} A.,   {Colacino} C.~N.,  2008, \mn@doi [\mnras]
  {10.1111/j.1365-2966.2008.13682.x}, \href
  {http://adsabs.harvard.edu/abs/2008MNRAS.390..192S} {390, 192}

\bibitem[\protect\citeauthoryear{{Sesana}, {Vecchio}  \& {Volonteri}}{{Sesana}
  et~al.}{2009}]{Sesana2009}
{Sesana} A.,  {Vecchio} A.,   {Volonteri} M.,  2009, \mn@doi [\mnras]
  {10.1111/j.1365-2966.2009.14499.x}, \href
  {http://adsabs.harvard.edu/abs/2009MNRAS.394.2255S} {394, 2255}

\bibitem[\protect\citeauthoryear{{Sesana}, {Gair}, {Berti}  \&
  {Volonteri}}{{Sesana} et~al.}{2011a}]{Sesana2011a}
{Sesana} A.,  {Gair} J.,  {Berti} E.,   {Volonteri} M.,  2011a, \mn@doi [\prd]
  {10.1103/PhysRevD.83.044036}, \href
  {http://adsabs.harvard.edu/abs/2011PhRvD..83d4036S} {83, 044036}

\bibitem[\protect\citeauthoryear{{Sesana}, {Gualandris}  \& {Dotti}}{{Sesana}
  et~al.}{2011b}]{Sesana2011b}
{Sesana} A.,  {Gualandris} A.,   {Dotti} M.,  2011b, \mn@doi [\mnras]
  {10.1111/j.1745-3933.2011.01073.x}, \href
  {http://adsabs.harvard.edu/abs/2011MNRAS.415L..35S} {415, L35}

\bibitem[\protect\citeauthoryear{{Sesana}, {Barausse}, {Dotti}  \&
  {Rossi}}{{Sesana} et~al.}{2014}]{Sesana2014}
{Sesana} A.,  {Barausse} E.,  {Dotti} M.,   {Rossi} E.~M.,  2014, \mn@doi
  [\apj] {10.1088/0004-637X/794/2/104}, \href
  {http://adsabs.harvard.edu/abs/2014ApJ...794..104S} {794, 104}

\bibitem[\protect\citeauthoryear{{Shakura} \& {Sunyaev}}{{Shakura} \&
  {Sunyaev}}{1973}]{Shakura1973}
{Shakura} N.~I.,  {Sunyaev} R.~A.,  1973, \aap, \href
  {http://adsabs.harvard.edu/abs/1973A%26A....24..337S} {24, 337}

\bibitem[\protect\citeauthoryear{{Shen}, {Mo}, {White}, {Blanton}, {Kauffmann},
  {Voges}, {Brinkmann}  \& {Csabai}}{{Shen} et~al.}{2003}]{Shen2003}
{Shen} S.,  {Mo} H.~J.,  {White} S.~D.~M.,  {Blanton} M.~R.,  {Kauffmann} G.,
  {Voges} W.,  {Brinkmann} J.,   {Csabai} I.,  2003, \mn@doi [\mnras]
  {10.1046/j.1365-8711.2003.06740.x}, \href
  {http://adsabs.harvard.edu/abs/2003MNRAS.343..978S} {343, 978}

\bibitem[\protect\citeauthoryear{{Shi}, {Krolik}, {Lubow}  \& {Hawley}}{{Shi}
  et~al.}{2012}]{Shi2012}
{Shi} J.-M.,  {Krolik} J.~H.,  {Lubow} S.~H.,   {Hawley} J.~F.,  2012, \mn@doi
  [\apj] {10.1088/0004-637X/749/2/118}, \href
  {http://adsabs.harvard.edu/abs/2012ApJ...749..118S} {749, 118}

\bibitem[\protect\citeauthoryear{{Shimizu}, {Kitayama}, {Sasaki}  \&
  {Suto}}{{Shimizu} et~al.}{2002}]{Shimizu2002}
{Shimizu} M.,  {Kitayama} T.,  {Sasaki} S.,   {Suto} Y.,  2002, \mn@doi [\pasj]
  {10.1093/pasj/54.5.645}, \href
  {http://adsabs.harvard.edu/abs/2002PASJ...54..645S} {54, 645}

\bibitem[\protect\citeauthoryear{{Springel}}{{Springel}}{2010}]{Springel2010}
{Springel} V.,  2010, \mn@doi [\mnras] {10.1111/j.1365-2966.2009.15715.x},
  \href {http://adsabs.harvard.edu/abs/2010MNRAS.401..791S} {401, 791}

\bibitem[\protect\citeauthoryear{{Taffoni}, {Mayer}, {Colpi}  \&
  {Governato}}{{Taffoni} et~al.}{2003}]{Taffoni2003}
{Taffoni} G.,  {Mayer} L.,  {Colpi} M.,   {Governato} F.,  2003, \mn@doi
  [\mnras] {10.1046/j.1365-8711.2003.06395.x}, \href
  {http://adsabs.harvard.edu/abs/2003MNRAS.341..434T} {341, 434}

\bibitem[\protect\citeauthoryear{{Tamanini}, {Caprini}, {Barausse}, {Sesana},
  {Klein}  \& {Petiteau}}{{Tamanini} et~al.}{2016}]{Tamanini2016}
{Tamanini} N.,  {Caprini} C.,  {Barausse} E.,  {Sesana} A.,  {Klein} A.,
  {Petiteau} A.,  2016, \mn@doi [\jcap] {10.1088/1475-7516/2016/04/002}, \href
  {http://adsabs.harvard.edu/abs/2016JCAP...04..002T} {4, 002}

\bibitem[\protect\citeauthoryear{{Tamburello}, {Capelo}, {Mayer}, {Bellovary}
  \& {Wadsley}}{{Tamburello} et~al.}{2017}]{Tamburello2017}
{Tamburello} V.,  {Capelo} P.~R.,  {Mayer} L.,  {Bellovary} J.~M.,   {Wadsley}
  J.~W.,  2017, \mn@doi [\mnras] {10.1093/mnras/stw2561}, \href
  {http://cdsads.u-strasbg.fr/abs/2017MNRAS.464.2952T} {464, 2952}

\bibitem[\protect\citeauthoryear{{Tang}, {Haiman}  \& {MacFadyen}}{{Tang}
  et~al.}{2018}]{Tang2018}
{Tang} Y.,  {Haiman} Z.,   {MacFadyen} A.,  2018, \mn@doi [\mnras]
  {10.1093/mnras/sty423}, \href
  {http://adsabs.harvard.edu/abs/2018MNRAS.476.2249T} {476, 2249}

\bibitem[\protect\citeauthoryear{{Tremmel}, {Karcher}, {Governato},
  {Volonteri}, {Quinn}, {Pontzen}, {Anderson}  \& {Bellovary}}{{Tremmel}
  et~al.}{2017}]{Tremmel2017}
{Tremmel} M.,  {Karcher} M.,  {Governato} F.,  {Volonteri} M.,  {Quinn} T.~R.,
  {Pontzen} A.,  {Anderson} L.,   {Bellovary} J.,  2017, \mn@doi [\mnras]
  {10.1093/mnras/stx1160}, \href
  {http://cdsads.u-strasbg.fr/abs/2017MNRAS.470.1121T} {470, 1121}

\bibitem[\protect\citeauthoryear{{Tremmel}, {Governato}, {Volonteri}, {Quinn}
  \& {Pontzen}}{{Tremmel} et~al.}{2018}]{Tremmel2018}
{Tremmel} M.,  {Governato} F.,  {Volonteri} M.,  {Quinn} T.~R.,   {Pontzen} A.,
   2018, \mn@doi [\mnras] {10.1093/mnras/sty139}, \href
  {http://adsabs.harvard.edu/abs/2018MNRAS.475.4967T} {475, 4967}

\bibitem[\protect\citeauthoryear{{Valiante}, {Schneider}, {Volonteri}  \&
  {Omukai}}{{Valiante} et~al.}{2016}]{Valiante2016}
{Valiante} R.,  {Schneider} R.,  {Volonteri} M.,   {Omukai} K.,  2016, \mn@doi
  [\mnras] {10.1093/mnras/stw225}, \href
  {http://adsabs.harvard.edu/abs/2016MNRAS.457.3356V} {457, 3356}

\bibitem[\protect\citeauthoryear{{Vasiliev}}{{Vasiliev}}{2014}]{Vasiliev2014}
{Vasiliev} E.,  2014, \mn@doi [Classical and Quantum Gravity]
  {10.1088/0264-9381/31/24/244002}, \href
  {http://adsabs.harvard.edu/abs/2014CQGra..31x4002V} {31, 244002}

\bibitem[\protect\citeauthoryear{{Vasiliev}}{{Vasiliev}}{2016}]{Vasiliev2016}
{Vasiliev} E.,  2016, in {Meiron} Y.,  {Li} S.,  {Liu} F.-K.,   {Spurzem} R.,
  eds,  IAU Symposium Vol. 312, Star Clusters and Black Holes in Galaxies
  across Cosmic Time. pp 92--100 (\mn@eprint {arXiv} {1411.1762}),
  \mn@doi{10.1017/S1743921315007607}

\bibitem[\protect\citeauthoryear{{Vasiliev}, {Antonini}  \&
  {Merritt}}{{Vasiliev} et~al.}{2015}]{Vasiliev2015}
{Vasiliev} E.,  {Antonini} F.,   {Merritt} D.,  2015, \mn@doi [\apj]
  {10.1088/0004-637X/810/1/49}, \href
  {http://adsabs.harvard.edu/abs/2015ApJ...810...49V} {810, 49}

\bibitem[\protect\citeauthoryear{{Vogelsberger} et~al.,}{{Vogelsberger}
  et~al.}{2014}]{Vogelsberger2014}
{Vogelsberger} M.,  et~al., 2014, \mn@doi [\nat] {10.1038/nature13316}, \href
  {http://adsabs.harvard.edu/abs/2014Natur.509..177V} {509, 177}

\bibitem[\protect\citeauthoryear{{Volonteri}, {Haardt}  \& {Madau}}{{Volonteri}
  et~al.}{2003}]{Volonteri2003}
{Volonteri} M.,  {Haardt} F.,   {Madau} P.,  2003, \mn@doi [\apj]
  {10.1086/344675}, \href {http://adsabs.harvard.edu/abs/2003ApJ...582..559V}
  {582, 559}

\bibitem[\protect\citeauthoryear{{Volonteri}, {Lodato}  \&
  {Natarajan}}{{Volonteri} et~al.}{2008}]{Volonteri2008}
{Volonteri} M.,  {Lodato} G.,   {Natarajan} P.,  2008, \mn@doi [\mnras]
  {10.1111/j.1365-2966.2007.12589.x}, \href
  {http://adsabs.harvard.edu/abs/2008MNRAS.383.1079V} {383, 1079}

\bibitem[\protect\citeauthoryear{{Wadsley}, {Keller}  \& {Quinn}}{{Wadsley}
  et~al.}{2017}]{Wadsley2017}
{Wadsley} J.~W.,  {Keller} B.~W.,   {Quinn} T.~R.,  2017, \mn@doi [\mnras]
  {10.1093/mnras/stx1643}, \href
  {http://adsabs.harvard.edu/abs/2017MNRAS.471.2357W} {471, 2357}

\bibitem[\protect\citeauthoryear{{White} \& {Rees}}{{White} \&
  {Rees}}{1978}]{White1978}
{White} S.~D.~M.,  {Rees} M.~J.,  1978, \mn@doi [\mnras]
  {10.1093/mnras/183.3.341}, \href
  {http://adsabs.harvard.edu/abs/1978MNRAS.183..341W} {183, 341}

\bibitem[\protect\citeauthoryear{{Wyithe} \& {Loeb}}{{Wyithe} \&
  {Loeb}}{2003}]{Wyithe2003}
{Wyithe} J.~S.~B.,  {Loeb} A.,  2003, \mn@doi [\apj] {10.1086/375187}, \href
  {http://adsabs.harvard.edu/abs/2003ApJ...590..691W} {590, 691}

\bibitem[\protect\citeauthoryear{{Yu}}{{Yu}}{2002}]{Yu2002}
{Yu} Q.,  2002, \mn@doi [\mnras] {10.1046/j.1365-8711.2002.05242.x}, \href
  {http://adsabs.harvard.edu/abs/2002MNRAS.331..935Y} {331, 935}

\bibitem[\protect\citeauthoryear{{del Valle} \& {Escala}}{{del Valle} \&
  {Escala}}{2012}]{delValle2012}
{del Valle} L.,  {Escala} A.,  2012, \mn@doi [\apj]
  {10.1088/0004-637X/761/1/31}, \href
  {http://adsabs.harvard.edu/abs/2012ApJ...761...31D} {761, 31}

\bibitem[\protect\citeauthoryear{{del Valle}, {Escala}, {Maureira-Fredes},
  {Molina}, {Cuadra}  \& {Amaro-Seoane}}{{del Valle}
  et~al.}{2015}]{DelValle2015}
{del Valle} L.,  {Escala} A.,  {Maureira-Fredes} C.,  {Molina} J.,  {Cuadra}
  J.,   {Amaro-Seoane} P.,  2015, \mn@doi [\apj] {10.1088/0004-637X/811/1/59},
  \href {http://cdsads.u-strasbg.fr/abs/2015ApJ...811...59D} {811, 59}

\makeatother
\end{thebibliography}

\end{document}